\documentclass[12pt,preprint]{aastex}
\usepackage{times}
\usepackage{enumerate}
\usepackage{graphicx}
\usepackage{natbib}
\usepackage{epstopdf}
\usepackage{rotating}
\newcommand{\ltsima}{\stackrel{\textstyle <}{\sim}}
\newcommand{\simlt}{\scriptsize{\raisebox{-2pt}{$\ltsima$}}\normalsize}
\newcommand{\gtsima}{\stackrel{\textstyle >}{\sim}}
\newcommand{\simgt}{\scriptsize{\raisebox{-2pt}{$\gtsima$}}\normalsize}

\renewcommand{\thepage}{\rm\arabic{page}}

\newcommand{\swas}{{\em SWAS}}
\newcommand{\hhifi}{{\em Herschel}/HIFI}

\newcommand{\water}{H$_2$O}

\newcommand{\nwater}{H$_2^{\:16}$O}

\newcommand{\oxy}{O$_2$}
\newcommand{\nco}{$^{12}$CO}
\newcommand{\ico}{$^{13}$CO}
\newcommand{\cio}{C$^{18}$O}
\newcommand{\cth}{C$_{2}$H}
\newcommand{\nthp}{N$_{2}$H$^{+}$}

\newcommand{\asec}{$^{\prime\prime}$}
\newcommand{\amin}{$^{\prime}$}
\newcommand{\ddeg}{$^{\rm o}$}
\newcommand{\mh}{H$_2$}

\newcommand{\av}{{\em A}$_{\rm V}$}
\newcommand{\um}{$\:\mu$m}
\newcommand{\kms}{~km~s$^{{-1}}$}
\newcommand{\cmc}{cm$^{-3}$}
\newcommand{\cms}{cm$^{-2}$} 

\newcommand{\lsol}{L$_{\odot}$}

\newcommand{\go}{$G_{\rm o}$}
\newcommand{\dash}{$\,$--$\,$}

\newcommand{\tro}{1$_{10}\,$--$\,$1$_{01}$}

\newcommand{\tgas}{$T_{\rm gas}$}

\newcommand{\etal}{et$\,$al.}
\newcommand{\ti}{$\,\times\,$}

\renewcommand{\apj}{{\em Ap.~J.}}
\renewcommand{\apjl}{{\em Ap.~J.~Letters}}
\renewcommand{\apjs}{{\em Ap.~J.~Suppl.}}
\renewcommand{\araa}{{\em A.R.A.\&A.}}

\renewcommand{\mnras}{{\em M.N.R.A.S.}}

\renewcommand{\aap}{{\em A\&A}}
\renewcommand{\aj}{{\em A.~J.}}

\makeatletter
\def\fnum@figure{{Fig.~\thefigure}}
\long\def\@makecaption#1#2{
 \vskip 10pt 
 \setbox\@tempboxa\hbox{#1. #2}
 \ifdim \wd\@tempboxa >\hsize \unhbox\@tempboxa\par \else \hbox
to\hsize{\hfil\box\@tempboxa\hfil} 
 \fi}

\makeatother

\makeatletter
\def\ps@myheadings{\let\@mkboth\@gobbletwo
\def\@oddhead{\hbox{}\sl\rightmark \hfil \thepage}%
\def\@oddfoot{}\def\@evenhead{\thepage\hfil\sl\leftmark\hbox {}}%
\def\@evenfoot{}\def\sectionmark##1{}\def\subsectionmark##1{}}
\makeatother

\makeatletter
\def\subsection{\@startsection{subsection}{2}{\z@}{-3.25ex plus -1ex minus 
   -.2ex}{1.5ex plus .2ex}{ \em \rm}}
\makeatother

\newcounter{ncount}

\shorttitle{Water Vapor is a Surface Tracer}
\shortauthors{Melnick et al.}

\begin{document}

\begin{center}
{\large\bf {Distribution of Water Vapor in Molecular Clouds. II}} \\*[14mm]
Gary J. Melnick$^1$, Volker Tolls$^1$, Ronald L. Snell$^2$, Michael J. Kaufman$^3$, Edwin A. Bergin$^4$,
Javier  R. Goicoechea$^5$, Paul F. Goldsmith$^6$, Eduardo Gonz\'{a}lez-Alfonso$^7$, David J. Hollenbach$^8$, 
Dariusz C. Lis,$^6$, and David A. Neufeld$^9$ \\
\vspace{34mm}

%Revised:~~February 3, 2020\\*[2mm]
%{\em \today} \\*[11mm]

\vspace{2.8in}

%Revision 2\\*[18mm]
Accepted to appear in the {\em Astrophysical Journal} February 17, 2020
%Received$\;$\rule{1.65in}{0.25mm}$\,$;~~~~~Accepted$\;$\rule{1.65in}{0.25mm}
\end{center}

\vspace{2.6in}

\clearpage

\setcounter{page}{2}

\pagestyle{myheadings}
\markright{{}\hfill{\hbox{\hss\rm Page }}}

\begin{enumerate}
\item Harvard-Smithsonian Center for Astrophysics, 60 Garden Street, MS 66, Cambridge, MA 02138, USA \\*[-6mm]

\item Department of Astronomy, University of Massachusetts, Amherst, MA 01003, USA \\*[-6mm]

\item Department of Physics and Astronomy, San Jos\'{e} State University, San Jose, CA 95192, USA \\*[-6mm]

\item Department of Astronomy, The University of Michigan, 500 Church Street, Ann Arbor, MI 48109-1042, USA \\*[-6mm]

\item IFF, Consejo Superior de Investigaciones Cientificas (CSIC), 28049 Madrid, Spain \\*[-6mm]

\item Jet Propulsion Laboratory, California Institute of Technology, 4800 Oak Grove Drive, Pasadena, CA 91109, USA \\*[-6mm]

\item Universidad de Alcal\'{a} de Henares, Departamento de F\'{i}sica, Campus Universitario, E-28871 Alcal\'{a} de Henares, Madrid, Spain \\*[-6mm]

\item SETI Institute, Mountain View, CA 94043, USA \\*[-6mm]

\item Department of Physics and Astronomy, Johns Hopkins University, 3400 North Charles Street, Baltimore, MD 21218, USA

\end{enumerate}

%\begin{center}
%ABSTRACT
%\end{center}

\renewcommand{\baselinestretch}{1.45}

\begin{abstract}
The depth-dependent abundance of both gas-phase and solid-state water within dense, quiescent molecular clouds 
is important to both the cloud chemistry and gas cooling.  Where water is in the gas phase, it's free to participate
in a network of ion-neutral reactions that lead to a host of oxygen-bearing molecules, and its many 
energy levels make it an effective coolant for gas temperatures $\geq\:$20$\:$K.  Where water is abundant 
as ice on grain surfaces, and unavailable to cool the gas, significant amounts of oxygen are removed from the 
gas phase, suppressing the gas-phase chemical reactions that lead to a number of oxygen-bearing species, including \oxy.
Models of FUV-illuminated clouds predict that the gas-phase water abundance peaks within \av$\,\sim\,$3 and 8 
mag.~of the cloud surface, depending on the gas density and FUV field strength.  Deeper within such 
clouds, water is predicted to exist mainly as ice on grain surfaces.  More broadly, these models are used to analyze a variety 
of other regions, including outflow cavities associated with young stellar objects and the surface layers of protoplanetary disks.
In this paper, we report the results of observational tests of FUV-illuminated cloud models toward the Orion Molecular Ridge 
and Cepheus B using data obtained from the {\em Herschel Space Observatory} and the Five College Radio Astronomy 
Observatory.  Toward Orion, 2,220 spatial positions were observed along the face-on Ridge in the \water\ \tro\ 557~GHz
and NH$_3$ $J, K =\:$1,0\dash 0,0 572~GHz lines.  Toward Cepheus B, two strip scans were made 
in the same lines across the edge-on ionization front.  These new observations demonstrate that gas-phase water exists
primarily within a few magnitudes of dense cloud surfaces, strengthening the conclusions of an earlier study based
on a much smaller dataset, and indirectly supports the prediction that water ice is quite abundant in
dense clouds.
\end{abstract}

\keywords{astrochemistry -- ISM: abundances -- ISM: individual objects (Orion) -- ISM: molecules -- submillimeter: ISM}

\renewcommand{\baselinestretch}{1.0}

\vspace{4.5in}

\baselineskip=24pt

\clearpage

\section{INTRODUCTION}

The distribution of gas-phase water within dense ($n$(\mh)$\:\simgt\:$10$^3\:$\cmc) molecular clouds is 
of interest since it affects both the oxygen chemistry and cooling within the gas.  In particular, where water is 
present in the gas phase, it is free
to participate in ion-neutral chemical reactions that affect the distribution of oxygen among
species such as O, OH, \oxy, and CO.  Where water is present as ice, which results from 
the freeze-out of gas-phase water onto dust grains or direct formation on grain surfaces via 
the repeated hydrogenation of O,
the $\simgt\:$90$\;$K sublimation temperature required to release water from
these grains effectively means that within cold (i.e., $\simlt\;$30$\:$K) dense clouds,
water, and the oxygen it contains, remains locked as ice and unavailable for further reactions in the
gas phase.

In an earlier study \citep[][henceforth Paper I]{Melnick11}, ground-state \tro\ ortho-\water\ 557~GHz 
observations obtained using the
{\em Submillimeter Wave Astropnomy Satellite} ({\em SWAS}) and millimeter-wave spectral line observations 
obtained using the Five College Radio Astronomy Observatory (FCRAO) were used to determine
the depth-dependent distribution of gas-phase water toward the face-on Orion Molecular Cloud ridge, located
at a distance of 412 pc \citep{Reid09}.  The present
work reexamines the results of this previous study, using instead \tro\ ortho-\water\ observations obtained
with the {\em Herschel Space Observatory}.  Because of its larger aperture (and smaller beam size) and 
greater sensitivity, {\em Herschel} was able to sample 2,220 spatial positions along the Orion ridge 
whereas \swas\ sampled only 77 spatial positions over approximately the same area.

As noted in the earlier study, establishing the distribution of gas-phase water within dense molecular clouds
is important for at least two reasons.  First, the depth-dependent abundance of gas-phase water is sensitive
to the efficiency of several key processes,
such as photodissociation, photodesorption, gas-phase reactions, gas-grain reactions,
and grain-surface reactions, most of which depend upon the gas density and far-ultraviolet (FUV) flux 
(6\,eV\,$<$\,$h\nu$\,$<$\,13.6\,eV).  The relative importance of these processes remain somewhat uncertain.
Second, because oxygen is the most abundant element after hydrogen and helium, the processes that control
the amount of oxygen locked in both water vapor and water ice determine the amount of
residual oxygen free to react with other species. In this way, the predicted abundance of a host of species 
that depend on the gas-phase oxygen-to-carbon or oxygen-to-nitrogen ratio hinges on 
knowledge of the main reservoirs of oxygen, such as gas-phase water and water ice.  In their study of the distribution
of gas- and ice-phase water within molecular clouds, \citet{Hollenbach09} found that, over a broad range of cloud densities,
i.e., $n =\:$10$^3$\dash 10$^5\:$\cmc, and external FUV field intensity, i.e., 1\dash 10$^3$ times the 
average local interstellar radiation field in the FUV band \citep{Habing68} -- gas-phase water remains largely
a cloud surface phenomenon.  This conclusion was supported by the earlier {\em SWAS} data (Paper~I); however,
the availability of a substantially improved {\em Herschel} dataset warrants a reexamination of the question.

In addition to Orion, we also present here results obtained toward the Cepheus B molecular cloud.  
The Cepheus B molecular cloud is at a distance of approximately 725 pc and
located adjacent to the Cepheus OB3 association.  The interface between
the molecular cloud and the OB association is clearly delineated by the
optically visible HII region S155.  S155 has a very sharp western edge
indicating the presence of an ionization front bounding the molecular cloud.
The ionizing radiation in this region is dominated by two hot and luminous 
stars: the O7 n star HD 217086 and the Be star HD 217061.  Whereas the photodissociation region (PDR) in Orion is viewed face-on,
the Cepheus B PDR displays an edge-on geometry and an opportunity to map the distribution of 
water through spatial strip scans.  

A number of previous investigations have invoked line-of-sight distributions of gas-phase and solid water for the purpose of
fitting observed line profiles \citep[e.g.,][]{Cernicharo06, Caselli12, Coutens12, Mottram13, Keto14, Schmalzl14}.  The 
results of these studies
are useful as they provide critical tests of chemical and dynamical models, often involving a mixture of quiescent envelopes, 
embedded heat sources and, in some cases, infall or outflow motions, along relatively
few sight lines.  However, whether it's line-of-sight complexity, the limited number of sight lines, or both, such observations
make it difficult to judge whether 
the predictions of chemical models of UV-illuminated dense, quiescent clouds and observations broadly agree.
The intent of this study is to utilize more than 2000 lines of sight to present a statistically robust observational study 
of the distribution of water vapor toward portions of the dense (i.e., $n$(\mh)$\:\simgt\,$10$^4$~\cmc) Orion 
Molecular Cloud ridge possessing no known embedded sources or outflows, and the fortuitous geometry of Cepheus B, to determine 
whether the presence of gas-phase water predominantly near molecular cloud surfaces is a ubiquitous property 
reflective of the processes outlined above.

In \S2 we discuss the observations.  In particular, in \S2.1, we discuss the observations toward
the Orion Molecular Cloud and, in \S2.2, we discuss the observations toward Cepheus~B.
In \S3 we present our findings.  The results toward Orion are given in \S3.1, including the {\em Herschel} 
maps obtained in \water\ 
and NH$_3$ (\S3.1.1); the determination of the visual extinction, \av, toward the Orion Ridge (\S3.1.2);  
the line optical depths, CO depletion, the relation between the \ico\ column density and \av\ (\S3.1.3); 
the depth-dependent line intensity ratios(\S3.1.4); a Principal Component Analysis (PCA)
of the lines observed (\S3.1.5); and, the H$_2$O Integrated Intensity vs. \av\ for a Range of \go\ 
and $n_{\rm H}$ (\S3.1.6).
In \S3.2 we present the results of the strip scans toward Cepheus B.  In \S4, we discuss the
implications of these results on our understanding of water in dense clouds and the interpretation of observatons of
gas-phase water.

\vspace{2mm}

\section{OBSERVATIONS}
\subsection{\em Orion Molecular Cloud}

The {\em Herschel}/HIFI observations of the Orion-KL region presented here were carried out on 2011 September 15 (ObsID: 1342228626). 
A 25\amin \ti 40\amin\ region was mapped in the 1$_{10}$ \dash 1$_{01}$ rotational transition of \water\
at 556.936 GHz in Band$\:$1b simultaneously in H- and V-polarization. The map was centered at 
$\alpha =\:$5$^{\rm h}$ 35$^{\rm m}$ 20.5$^{\rm s}$ and $\delta = -$5$^{\rm o}$ 17\amin\ 7\asec.  
HIFI \citep{Roelfsema12} was used in On-The-Fly (OTF) Maps observing mode with position-switch 
reference located at $\alpha =\,$5$^{\rm h}$ 32$^{\rm m}$ 25.9$^{\rm s}$ 
and $\delta = -$5$^{\rm o}$ 22\amin\ 36.80\asec\ (J2000).  The angular separation between adjacent scans was $\sim\,$40\asec, 
a compromise between area covered and spatial resolution. The HIFI receivers are sensitive to signals in both sidebands.  
Thus, the local oscillator (LO) frequency could be selected such that the (J,K)$\:=\:$(1,0)\dash (0,0) 
transition of ortho-NH$_3$ at 572.498 GHz could be observed simultaneously with \water. 
The wide band spectrometer (WBS) provided a 
spectral resolution of 1.1 MHz corresponding to a velocity width of $\sim\,$0.6\kms\ for the 556.9 GHz line.  
The total on-source integration time per map point (for each polarization) 
was 5.92 sec.~and the total mapping time was 5.1 hours. The {\em Herschel} telescope beam FWHM is
$\sim\,$37\asec\ at 557 GHz and 572 GHz, the main beam efficiency  \citep{Mueller14} is 0.62, 
and the ratio of Lower Sideband (LSB) to Upper Sideband (USB) gain is 0.54:0.46 at these frequencies \citep{Kester17}.
All values were extracted from the Observation Context HIFI Calibration Data Set (HIFI\_CAL\_25\_0).

All HIFI data were processed using the {\em Herschel} Interactive Processing Environment (HIPE) \citep{Ott10}, 
Version 10.3, up to Level 2 providing fully calibrated spectra with the intensities expressed as antenna temperature 
and the frequencies in the frame of the Local Standard of Rest (LSR). Further data processing steps in HIPE 
included running FitHifiFringeTask and applying the main beam efficiency before saving the data in 
CLASS-FITS format. The overall data quality was excellent with only minor low-intensity ripples in some scans. 

Since the initial extensive analysis, the data reduction software HIPE has been upgraded to version 15.0.  
In order to avoid the need to redo the data reduction and analysis, we updated our data sets to bring them into 
conformity with data reduced in HIPE, Version 15.0. These updates included changing the receiver sideband ratios and the main beam efficiency
to the values listed above and comparing random spectra (with line emission) reduced in HIPE Version 10.3 with 
the same spectra reduced in HIPE 15.0. The observed differences were sufficiently small (i.e., $<\,$3\%) so as to have no impact 
on the analysis.

In 2004, maps of the emission from \cth, HCN, \nthp, 
and CN toward Orion were obtained with the 14-meter telescope of the Five College Radio Astronomy Observatory (FCRAO). 
In April 2005, further observations were obtained in \cth\ and \nthp\ repeating regions in Orion where 
the emission was weak. All lines were observed using the On-The-Fly observing method. The 32-pixel SEQUOIA 
array receiver (Erickson et al. 1999) was used to observe two lines simultaneously. The spectrometer for each 
spatial pixel was a digital autocorrelator with a bandwidth of 50 MHz and 1024 spectral channels per pixel 
leading to a velocity channel spacing that varied between 0.13 and 0.17\kms\ depending on the line frequency. 
The observations were first reduced to form maps with data spaced by 20\asec.  The FWHM beam size of the 
FCRAO telescope varies from approximately 46\asec\ at the CN line frequency to 56\asec\ at   
the \cth\ line frequency.  The main beam efficiency, $\eta_{\rm mb}$, of the FCRAO antenna varies from 
approximately 0.45 (at 115 GHz) to 0.48 (at $\sim\,$88 GHz).  

The \nco\ and \ico\ data used for this analysis are the same as described in \citet{Ripple13}, 
who also used the FCRAO 32-pixel SEQUOIA array receiver during 2005 and 2006.  Both lines were observed
simultaneously using the On-The-Fly observing method. However, they used the autocorrelator 
with 25 MHz bandwidth and a spectral channel spacing of 0.077\kms\ and 0.080\kms\ at the
\nco\ and \ico\ line frequencies, respectively.  For the analysis, the spectral channels were resampled
to a spacing of 0.2\kms.  The FWHM beam sizes of the FCRAO telescope are approximately 45\asec\ 
and 47\asec\ and the main beam efficiencies are 0.45 and 0.48 at the \nco\ and \ico\ line frequencies, respectively. 
For more details see \citet{Ripple13}.  A summary of the \hhifi\ and FCRAO observations
is provided in Table~1.  The maps obtained with the FCRAO telescope 
are shown in Fig.~\ref{fcraomaps}.

In order to directly compare the data obtained from \hhifi\ and FCRAO, all
data were first regridded onto a common spatial grid with equal beam sizes. 
Since the area covered by the FCRAO maps exceeds those of the \hhifi\ observations, 
the latter determined the extent of the final maps.  The regridded maps have 24 $\times$ 40 spatial pixels
on a 1\amin\ grid and beam sizes that were convolved to a FWHM of 1\amin, slightly larger 
than the largest beam of the observations (see Table~1), using appropriate Gaussian kernels.
The spatial positions of the regridded data are shown in Fig.~\ref{regridpos}.

The integrated intensities were derived in two steps.  First, Gaussian fits to the 
continuum-subtracted line profiles of all the species 
listed in Table~1 were obtained.  In many cases, these fits involved multiple velocity components including,
for HCN, CN, C$_2$H, and N$_2$H$^+$, hyperfine components of known velocity separation (from
the main component) and common FWHM (to the main component).  

In the second step, a direct, summed integrated intensity was obtained for each molecule and map position.  To do so, 
the fitted line centers and FWHM's of each \ico\ velocity component were used to set the velocity 
ranges within each spectrum in which the flux in each spectral resolution element 
was directly summed.  Specifically, the region within each spectrum directly summed was bounded by
the \ico\ component with the lowest velocity, 
minus 1.3 times the FWHM of this component, to the \ico\ component with the highest velocity, plus 1.3 times its FWHM. 
Thus, the data set for further analysis included the fit results and integrated intensities of all 
individual \ico\ velocity components together
with the summed integrated intensities over the full velocity range, as described above, for all other species.
The first step was necessitated by the desire to derive
line-center optical depths along with the total column density of \ico\ toward each spatial position 
\citep[see][eqns.~(1)\dash (5)]{Ripple13}, as well as to
examine the connection between the LSR velocities of all the observed species.  The second step, i.e.,
direct integration of the spectra to determine the integrated intensities, was necessitated by the
desire to include the flux contribution from noisy, but net positive, spectral lines whose Gaussian fits
were either difficult or impossible to obtain.  As will be discussed, species whose abundance (and emission)
peak within a relatively narrow range of \av's will produce weak lines toward lines of sight with 
lower \mh\ densities.  Nevertheless, such lines contain useful information that can be
retrieved through direct integration.

\vspace{2mm}

\subsection{{\em Cepheus B Molecular Cloud}}

The {\em Herschel}/HIFI observations of Cepheus B were carried out on 2012 May 8 and 
2012 May 24. Observations of the 557 GHz line of water were obtained in two strip scans using
the OTF mapping mode of {\em Herschel}.   A northern strip map (ObsIDs: 1342245588) was centered at
RA\,(J2000) = 22$^{\rm h}$ 57$^{\rm m}$ 16$^{\rm s}$ and Dec\,(J2000) = +62$^{\rm o}$ 35\amin\ 45\asec\ 
and covered a range in right ascension from 193.3 arcseconds west to 189.0 arcseconds east of the 
center.  A southern strip map (ObsID: 1342246073) was centered at RA\,(J2000) = 22$^{\rm h}$ 57$^{\rm m}$ 24$^{\rm s}$ 
and Dec\,(J2000) = +62$^{\rm o}$ 34\amin\ 45\asec\ and covered a range in right ascension from 192.9 
arcseconds west to 190.1 arcseconds east of the center. The HIFI receiver setup for both observations was the 
same as for the Orion-KL observations described above.

Observations of the 572 GHz 
line of NH$_3$ were obtained simulataneously with those of \water\ in the other sideband of the water
line. These data were reduced using HIPE in the same manner as described in \S2.1.
The two strip maps were scanned along an east-west line and included 
81 samples spaced by 5\asec, the two strip maps were offset in 
declination by one arcminute.   The location of the
strip maps are shown in Fig.~\ref{cephscanfig} superimposed on an image of the region. 

The data from the H- and V-beams, separated by $\sim\,$4\asec\ orthogonal to 
the OTF scan direction were coadded.  Nevertheless, the signal 
to noise in the coadded spectra was still relatively low.  We therefore
averaged the spectra along each strip map in groups of seven spectra (with equal
weighting) forming a new spectrum every fourth spectrum.  The resulting 
new strip maps have 19 spectra along the northern scan and 20 spectra along the southern scan
spaced by 20\asec\ and averaged over 30\asec\
(approximately the {\em Herschel} beam at 557 GHz and much larger than 
the H- and V-beam separation).  In Fig.~\ref{cephspectranorth}, we show plots of the resulting
integrated intensity of the water and ammonia lines as a function of offset 
from the center position for the northern strip map, while Fig.~\ref{cephspectrasouth} shows
the same for the southern strip map.  The
integration was performed over a velocity range of -20 to -10 \kms, except for a few spectra
that showed broad wings, in which the velocity range was extended to -30 to
0 \kms. 

In the course of making the strip maps, a broad water emission source was discovered in 
several of the spectra in both strip maps.  Fig.~\ref{cephoutflow} shows the water spectrum 
toward one of these positions.  This previously unreported outflow is discussed in \S 3.2.

\vspace{2mm}

\section{RESULTS}
\subsection{{\em Orion Molecular Cloud}}

\subsubsection{{\it Water and Ammonia Maps}}

Fig.~\ref{h2omap} shows the \hhifi\ ortho-\water\ 556.936 GHz \tro\ summed integrated intensity map
toward the Orion Molecular Ridge.  Fig.~\ref{nh3map} shows the \hhifi\ ortho-NH$_3$ 572.498 GHz 
(J,K)$\,=\,$(1,0)\dash (0,0) summed integrated intensity map toward the same region.

\subsubsection{\it \av\ {\it Determination}}

In Paper I, the relation between the \cio\ column density, $N$(\cio), and \av\ was based on studies of dark clouds, such as
Chamaeleon I and III-B \citep{Kainulainen06} using Two-Micron All Sky Survey (2MASS) data.
More recently, \citet{Ripple13} studied the relationship between $N$(\ico) and visual extinction toward the 
Orion A and Orion B molecular clouds.  They computed the \ico\ column density in these clouds from 
observations of the \nco\ $J = $1\dash 0 and \ico\ $J = $1\dash 0 transitions assuming LTE.  The visual extinction 
was determined directly from photometry of background stars using the 2MASS
database and used to establish the correlation between $N$(\ico) and \av. 
They find three distinct regimes of \av\ in which the ratio of \ico\ column density and visual extinction 
differ.  At low \av's (Regime 1), in the photon-dominated envelope, the column density of \ico\ is low, but 
increases into the well-shielded interior of the clouds at higher \av's (Regime 2).  At the highest \av's (Regime 3), some 
regions of the cloud, particularly those with dust temperatures below $\sim$\,22\dash 25\,K, 
show evidence of a flattening of the \ico\ integrated intensity, which is
attributed to CO depletion.

Based on our determination of $N$(\ico), we estimate the visual extinction using the results from 
\citet{Ripple13}. 
In the Ripple et al.~study, the Orion A cloud is divided into a number of partitions, where each partition
corresponds to a distinct spatial region within the larger Orion complex.  Partition 6 of their 
study corresponds to the region of the Orion A cloud investigated in this paper.  Within this partition,  
Ripple et al.~find no evidence for CO depletion.  This finding is in agreement with the conclusion
drawn in Paper~I.

The relation between $N$(\ico) and \av\ for Partition 6 is given by Ripple et al.~as:

\vspace{-7.5mm}

\begin{equation}
A_V ({\rm mag.}) =~2.0 \times 10^{-15}~N({\rm ^{13}CO})\,-\,0.5~~~~{\rm for}~2.5 \times 10 ^{14}~{\rm cm^{-2}} < N({\rm ^{13}CO}) < 2.1 \times 10^{15}~{\rm cm^{-2}} \\*[3mm]
\end{equation}
\begin{equation}
A_V ({\rm mag.}) =~2.78 \times 10^{-16}~N({\rm ^{13}CO})\,+\,3.12~~~~~{\rm for}~N({\rm ^{13}CO}) > 2.1 \times 10^{15}~{\rm cm^{-2}}\hspace{10mm}\phantom{0}
\end{equation}

\vspace{1mm}

Within the low column density regime, as expressed in Eqn.~(1), $N$(\ico) grows very slowly with increasing \av, 
consistent with the photodestruction of \ico\ near the cloud surface.  Ripple et al.~derived Eqn.~(1) based on an 
empirical fit to $N$(\ico) and \av\ data and we note that values of 
$N$(\ico)$\:<\:$2.5\ti 10$^{14}$~\cms\ would imply an unphysical value for \av.  However, of the 898 map 
positions used to construct the water map shown in Fig.~\ref{h2omap}, only two positions have $N$(\ico) values less than 
2.5 \ti 10$^{14}$~\cms.  These positions are located at the east edge of the map and have no associated 
\water\ emission and, thus, do not factor into the analysis that follows.  In fact, all but four map positions
have $N$(\ico)$ \;>\;$4\ti 10$^{14}$~\cms\ resulting in values of \av$\,>\:$0.3 mag. 

As the depth into the cloud increases, as expressed in Eqn.~(2),
and self shielding becomes effective, $N$(\ico) exhibits a steeper linear growth with \av.  Ripple et al.~find that in 
Partition 6 (Orion A), a significant fraction of \ico\ remains in the gas phase to at least \av\,$=\:$25 mag.  Fig.~\ref{avmap} shows
the map of \av\ generated using Eqns.~(1) and (2) and the \ico\ column densities derived from the FCRAO data obtained here.

\vspace{2mm}

\subsubsection{{\it Line Optical Depths}}

The analyses in the following sections assume that the line-center optical depths are less than 1.  Here we show this to be
the case for \ico, \cio, HCN, \nthp, CN, and \cth.

The line-center optical depth of the \ico\,(1\dash 0) transition can be expressed as:

\vspace{-0.8mm}

\begin{equation}
\tau(^{13}{\rm CO}) = -{\rm ln} \left[ 1 - \frac{T_{\rm max}(^{13}{\rm CO}) / 5.289\:{\rm K}}
{1 / \left(e^{\,5.289\,{\rm K}/T_{\rm ex} (^{13}{\rm CO})} - 1\right) - 0.16}\right]~,
\end{equation}

\vspace{4.9mm}

where 5.289$\:$K $\equiv h\nu$\,(\ico)$/k$, with $\nu$\,(\ico)$\:=\:$110.201~GHz, the rest frequency of the 
\ico\,(1\dash 0) transition, $T_{\rm max\,}$(\ico) is the main beam brightness temperature at the peak of \ico\
line  \citep[cf.][]{Ripple13}, and $T_{\rm ex}$(\ico) is the \ico\ excitation temperature.  
As discussed in Section 3.1.6, the \mh\ densities required to account for the measured
\water\ line emission vary between about 3\ti 10$^4$~\cmc\ and 10$^5$~\cmc.  Based on a grid of RADEX models
within this range of densities, gas temperatures between 20$\:$K and 35$\:$K, and column densities corresponding 
to \av's between 3 and 30, the average excitation temperature of the
\ico\,(1\dash 0) line is the same as that of the \nco\,(1\dash 0) line to within about
3\%, and $T_{\rm ex}$ is thus given by

\vspace{-3.3mm}

\begin{equation}
T_{\rm ex} = \frac{5.532~{\rm K}}{{\rm ln} \left[1 + 5.532\:{\rm K} / \left(T_{\rm max}(^{12}{\rm CO}) + 0.8363\:{\rm K}\right) \right]}
\end{equation}

\vspace{5.3mm}

where 5.532~K $\,\equiv\,h\nu$(\nco)$/k$, with $\nu$(\nco)$\:=\:$115.271~GHz, the rest frequency of the 
\nco\,(1\dash 0) line. Based on the \nco\ and \ico\ FCRAO measurements described in \S2.1, the \ico\ line-center 
optical depths are almost all
less than $\tau\,=\,$1.  This is illustrated in Figs.~\ref{13cotaumap} and \ref{13cotauplot}.

The molecules HCN, \nthp, CN, and \cth\ have hyperfine structure that can be used to estimate the 
optical depth of these lines. We fitted the hyperfine components in these spectra using a template that fixes the 
relative velocities of the hyperfine components, assumes a common line width and Gaussian line shape.  Thus, 
the free parameters for these fits are the intensities of the individual hyperfine components, one line velocity and 
one line width. These fits are then used to determine the intensity ratios of the various hyperfine components.

The hyperfine intensity ratio most sensitive to the optical depth is the ratio of the weakest hyperfine 
component to the strongest. For example, HCN has quadrupole hyperfine structure due to the spin of the 
nitrogen atom. The $J = $1\dash 0 transition of HCN is split into three components, 
the F=2-1 feature is the strongest and the F=0-1 feature is the weakest. If these components are 
optically thin and in LTE, the expected intensity ratio is 0.20.  In the upper-left panel of Fig.~\ref{hfsratios}
we plot the F=0-1/F=2-1 intensity ratio versus the intensity of the strongest F=2-1 hyperfine 
component for HCN, where the error bars indicate the 1$\sigma$ uncertainty in the intensity ratio. 
Although there is considerable scatter and uncertainty in the ratio for the weaker intensity 
lines, as the lines become stronger, the ratio approaches the LTE optically thin value of 0.2. We 
would expect the strongest lines to be the most optically thick; however, they are the lines with 
intensity ratios consistent with the optically thin LTE ratio.  Nevertheless, HCN has long been known to exhibit 
anomalous hyperfine ratios, as illustrated in the papers by \citet{Pirogov99} and 
\citet{Loughnane12}.  Hyperfine ratios that include the F=1-1 component are often the most anomalous, 
while the F=0-1/F=2-1 ratio is usually more consistent with its LTE value, especially in higher mass molecular 
cloud cores.  Despite this concern, the fact that the hyperfine ratios within the region studied 
here are consistent with an LTE ratio suggests that the HCN emission is optically thin.

\citet{Schilke92} observed the isotopomer H$^{13}$CN in several positions along the ridge in Orion,
though all of their observed positions lie within the denser gas region close to BN/KL excluded from our analysis 
(see Fig.~\ref{regridpos}).  Nevertheless,
they found the isotopic line ratio HCN/H$^{13}$CN to be of order 10, implying that the main isotopomer 
was optically thick. However, the intensity pattern of the hyperfine features in both isotopomers was roughly 
the same, which would be difficult to explain if HCN was optically thick.  In fact, the modeling by 
\citet{Mullins16} finds that, due to radiative transfer effects, the F$=$0-1 feature always appears stronger 
relative to the F$=$2-1 than predicted by LTE, just the opposite of what would be needed if HCN is 
optically thick.  Schilke et al.~suggest that fractionation of $^{13}$C is producing an increase in the 
$^{13}$C/$^{12}$C ratio in HCN. Thus, the low observed isotopic ratio could be either due to HCN 
being optically thick, or due to fractionation.  Since the hyperfine ratios could not be easily explained 
if HCN is optically thick, we believe that HCN is optically thin and fractionation is responsible for the 
small isotopic ratio.

The \nthp\ molecule also has quadrupole hyperfine structure produced by both nitrogen atoms.  The 
outermost nitrogen atom produces three hyperfine components (denoted by the quantum number F1) that are 
very similar to that in HCN, while the innermost nitrogen atom is responsible for further splitting (denoted 
by the quantum number F) in two of these three components \citep{Womack92}.
However, the splitting produced by the innermost nitrogen atom is much less than the line width of the emission in Orion. 
Thus, these hyperfine features are blended, resulting in only three resolvable \nthp\ hyperfine components. 
The line strength ratios of the resolvable features are the same as those in HCN. In the lower-left panel of 
Fig.~\ref{hfsratios}, we show the same hyperfine component ratios for \nthp, i.e., F1=0-1/F1=2-1, as 
discussed above for HCN. The result is similar, except in the case of \nthp, the intensity ratio approaches 
a value slightly greater than the optically thin LTE value of 0.2.  As with HCN, anomalous hyperfine ratios 
have been found in \nthp, and the study by \citet{Pirogov03} of the hyperfine ratios in the $J = $1\dash 0 
transition of \nthp\ in massive molecular cloud cores obtained results similar to those we find in Orion. 
We also find that the F=1-1/F=2-1 intensity ratio is slightly lower than the optically thin LTE ratio of 0.6, 
as did \citet{Pirogov03} in their study.  The origin of these hyperfine anomalies in \nthp\ and HCN has 
been investigated by \citet{Keto10}.  Although we find small hyperfine anomalies in \nthp, 
the F=0-1/F=2-1 intensity ratio still strongly suggests that the lines of \nthp, like HCN, are optically thin.

In CN, we observe the F=3/2-3/2, F=1/2-1/2, F=5/2-3/2, and F=3/2-1/2 hyperfine components of the $N = $1\dash 0 
transition.  The ratio of the weakest components (F=3/2-3/2 and F=1/2-1/2) to the strongest component (F=5/2-3/2)
has an optically thin LTE ratio of 0.30.  In the upper-right panel of 
Fig.~\ref{hfsratios} we plot the F=3/2-3/2/F=5/2-3/2 ratio versus the intensity of 
the F=5/2-3/2 component. As in the case for HCN, the data are consistent 
with the optically thin LTE ratio of 0.30.  A plot using the F=1/2-1/2/F=5/2-3/2 
intensity ratio is nearly identical to that shown.  These results indicate
that the CN emission is optically thin.

Finally, for the $N\!=\,$1\dash 0 transition of \cth, we observed the F=1-1 and F=0-1 
hyperfine components. The optically thin LTE intensity ratio of F=0-1/F=1-1 is 0.4.  The lower-right panel of 
Fig.~\ref{hfsratios} shows this ratio plotted versus the stronger F=1-1 component intensity.  As with the other 
molecules, the ratio indicates that \cth\ is also optically thin.

The emission from both \water\ and NH$_3$ is very weak. The observed transitions of these species have large 
critical densities; based on the molecular data in the Leiden Atomic and Molecular Database, 
the critical density for \water\ is 1\ti 10$^8$ \cmc\ at 25$\:$K, while that of NH$_3$ is
3 \ti 10$^7$ \cmc. In both cases, the critical density for these transitions is much larger than the gas density expected 
in this extended region of Orion.  \citet{Snell00} argued that the emission in such
high critical density transitions should be ``effectively'' optically thin, i.e., despite the actually 
optical depth of these lines, the emission should increase linearly with column density.

Although it is unlikely that water line photons are lost through collisional de-excitation, these photons 
can be repeatedly absorbed and reemitted before escaping. The large number of ``scattering'' events could 
alter the spatial distribution of the water emission, which would have to be accounted for in our analysis 
of the water map.  Whether spatial redistribution is important depends on the intrinsic line width of the 
water emission and details of the velocity field present within the molecular cloud.  However, scattering 
not only produces a redistribution of the emission spatially, it also redistributes the escaping photons 
in frequency away from the line center.  The classical consequence of a large number of scatterings is a 
double-horned line profile.   In Fig.~\ref{h2ospectra} we show examples of the water line profiles for a portion 
of the Orion Ridge. In all cases shown, and for spectra not shown, the line profiles are singly peaked and well 
fit by a Gaussian line shape. The absence of distorted line profiles argues strongly against any spatial 
redistribution of the water emission.

\vspace{2mm}

\subsubsection{\it Depth-Dependent Line Intensity Ratios}

Knowledge of the optically thin integrated intensity of each species coupled with the visual extinction 
for hundreds of lines of sight makes it possible to examine the intensity ratio of key species as a function of varying 
depth into the cloud.  Since this study is focused on the distribution of gas-phase water and the other species within dense 
quiescent gas, unaffected by the chemical and excitation changes induced by shocks, we restrict our analysis to 
the positions outside of the BN/KL and OMC-2 regions.  Excluded positions were determined based on
the measured line widths, i.e., positions having emission lines with widths greater than $\sim\,$10\kms\
were assumed to be indicative of non-quiescent material.  The regions included are shown in 
Fig.~\ref{regridpos}, panel (d).  In addition, we restrict our analysis to lines of sight having an \av\
between 0 and 30 magnitudes since the number of positions possessing an \av\ greater than 
30 magnitudes (beyond the regions already excluded) is limited (see Fig.~\ref{avmap}).
%and the unknown strength of the FUV field illuminating the far surface of the molecular ridge 
%(i.e., opposite to the face-on surface viewed from earth) is an additional uncertainty.  
A total of 636 lines of sight are thus considered in this analysis.

The depth-dependent emission of \ico, \cio, CN, HCN, C$_2$H, and N$_2$H$^+$,
which are based on FCRAO data, along with \water\ data obtained by {\em SWAS} was presented in Paper~I
(see Paper~I, Figs.$\:$11 and 12).  Fig.~\ref{ratiofig} here updates these earlier plots in three important ways:
(1) the use of an independent set of \water\ and NH$_3$ data obtained using {\em Herschel}; (2) the use of a 
significantly larger number of spatial positions sampled (636 versus 77 positions) obtained with the greater sensitivity 
and smaller beam size of {\em Herschel}; and (3) the use of the improved 
relation between the measured \ico\ column density and \av\ derived from the study by \citet{Ripple13}.

To reduce scatter in the 636 data points considered and, thus, better reveal any trends, 
the integrated intensity ratios have been co-averaged in bins of fixed $\,\delta$\av.  For each species, the number of 
\av\ bins needed to adequately reduce the $y$-axis scatter in each bin was determined from the total number of spectra
having a baseline signal-to-noise ratio $\geq\:$3.  As shown in Fig.~\ref{ratiofig}, between 7 and 13 \av\ bins were used
(with the greater number of \av\ bins corresponding to the species exhibiting the stronger, higher signal-to-noise emission
in need of less co-averaging).

The plotted $y$ value within each bin is the weighted mean,
$\mu\,=\,\Sigma(y_i/\sigma_i^2)/\Sigma(1/\sigma_i^2)$,
of the $i$ data points lying within that bin, 
where $y_i$ is the ratio of the integrated intensity, $I$, of species $a$ to species $b$ for point $i$,
i.e., $y_i=I_{a, i}/I_{b, i}$, and 
$\sigma_i\,=\,\left[y_i^{\;2}\,\left(\sigma_{a,i}^{\;2}/I_{a, i}^2\,+\,\sigma_{b,i}^{\;2}/I_{b, i}^2\right)\right]^{1/2}$,
where $\sigma_{a,i}$ and $\sigma_{b,i}$ are the 1$\sigma$ uncertainties associated with
the $i$th integrated intensity for species $a$ and $b$, respectively.
The 1$\sigma$ $y$-value error bars represent the uncertainty of the mean,
$\sigma_{\mu}\,=\,\left[1/\Sigma(1/\sigma_i^2)\right]^{1/2}$.  
%Though barely visible in these plots, the 1$\sigma$ error bars representing the $x$-value dispersion within each bin 
%are also shown.

Several things are evident from these plots.  First, as noted in Paper I, the \ico/\cio\ ratio profile, shown in panel (a) of 
Fig.~\ref{ratiofig}, indicates that the \ico\ emission exceeds that of \cio\ near the cloud surface, as expected
given \ico's higher abundance and, consequently, greater column density at a given \av, but declines as
both molecules become fully self-shielded, converging to
a value of between 8 and 12 deep in the cloud.  As explained in Paper I, this ratio is consistent with the measured 
$^{16}$O/$^{18}$O isotopic ratio of 500 and $^{12}$C/$^{13}$C ratio of between 43 
\citep{Hawkins87,Stacey93,Savage02} and 65 \citep{Langer90} measured toward Orion and suggests that such ratio
plots convey a correct picture of the abundance profiles.  However, since the Orion Molecular Ridge is not only
illuminated on its front, Earth-facing side, but almost certainly also on its far side by FUV radiation of unknown intensity,
the $x$-axis \av\ value at which intensity ratios peak carries some uncertainty.  Nonetheless, the relative intensity
behavior of the various species remains revealing.

Second, as discussed in Paper I (Paper I, Figs.$\:$15 and 16, and references therein), PDR models that include 
CN, HCN, and C$_2$H predict that these species exhibit their peak depth-dependent abundance at 
0$\;\leq\:$\av$\:\leq\:$10.  Fig.~\ref{ratiofig}, panels (c) \dash (e), show that these species peak more 
toward the cloud surface than throughout the cloud volume, in agreement with the model predictions.

Third, the increase in the gas-phase water abundance toward the cloud surface suggested in Paper~I is now clearly evident
in the larger data sample here (Fig.~\ref{ratiofig}, panel b).  Also, the rise in the N$_2$H$^+$/\cio\ ratio 
(Fig.~\ref{ratiofig}, panel g) beyond an \av\ of 10 is 
consistent with a decrease in the fractional ionization deep in the cloud as N$_2$H$^+$ is effectively removed by
N$_2$H$^+$ $+$ e$^-$ dissociative recombinations.  (The rise in this ratio at \av$\,<\,$10 likely results from the drop
in the \cio\ column density due to decreased self shielding.)

Can the peak in the gas-phase water abundance near an \av\ of 4 be no more than an excitation effect?  Specifically, 
if gas-phase water were distributed uniformly throughout the depth of the cloud, can an elevated gas temperature within
a few magnitudes of the cloud surface reproduce the \water\ profile shown in Fig.~\ref{ratiofig} (panel b)?  Gas temperatures
deep within the quiescent ridge have been estimated to be $\sim\,$25\dash 30$\:$K \citep{Ungerechts97}.  
At an \av\ of 4, the external FUV field strength is attenuated by more than a factor of 10$^3$.  Nevertheless, the field strength
toward the Orion Ridge remains sufficiently strong toward most lines of sight to warm the gas at an \av\ of  4 above 
that deeper into the cloud.  

In Paper I, the strength of the FUV field along the Orion Ridge was estimated based on 
[C\,II] 158\um\ observations obtained from the {\em Kuiper Airborne Observatory} with a 55\asec\ (FWHM) beam and a velocity
resolution of 67\kms\ \citep{Stacey93}.  More recently, \citet[][in preparation]{Pabst19} have undertaken a study of gas and dust tracers
in Orion A using {\em Herschel}, {\em Spitzer}, and {\em SOFIA} data.  In particular, use is made of a fully sampled,
1.2~deg$^2$ velocity-resolved [C\,II] map of Orion obtained with the upGREAT instrument onboard {\em SOFIA} with a 
15\asec\ (FWHM) beam.  The best fit to their data suggest values of \go\ that range between $\sim\,$5100 and 
$\sim\,$100 at 4\amin\ to 25\amin\ from $\Theta^1$C, respectively, with an average \go\ of approximately 500
over the area of the Ridge observed here.  The average density of
the Ridge is between $\sim\,$3\ti 10$^4$~\cmc\ and a few\ti 10$^5$~\cmc\
(see \citeauthor{Bally87} \citeyear{Bally87}; \citeauthor{Dutrey91} \citeyear{Dutrey91};
\citeauthor{Tatematsu93} \citeyear{Tatematsu93};
\citeauthor{Bergin94} \citeyear{Bergin94}; \citeauthor{Bergin96} \citeyear{Bergin96};
\citeauthor{Ungerechts97} \citeyear{Ungerechts97}; 
\citeauthor{Johnstone99} \citeyear{Johnstone99}).

\citet{Hollenbach09} have calculated the average gas temperature at which the water abundance is predicted to 
peak for a range of densities and FUV field strengths.  These results are shown in Fig.~\ref{gastemp} and indicate that
the gas temperature at the location of the peak water abundance ranges between about 25$\:$K and 30$\:$K for almost all
lines of site observed.  This result is consistent with 
a number of PDR models that have been developed in recent years that bracket the range of FUV intensities and densities
inferred for the Ridge \citep[e.g.,][]{Rollig07, Bisbas12, Lee14}.
In particular, these models consider a number of specific benchmark cases, including a gas density,
$n(= n_{\rm H} + 2n_{\rm H_2})$, of 3.2\ti 10$^5$~\cmc\ subject to a FUV field strength
10$^5$ Draine units ($=\:$1.7\ti Habing units, \go).  
Common to all these models is the prediction that the gas temperature is $\,\simlt\:\,$45$\:$K at \av$\,=\:$4 for  
\go$\,=\:$6\ti 10$^4$ and $n$(\mh)$\:=\:$1.6\ti 10$^5\:$\cmc\ (H being all molecular at this depth).  This is in good
agreement with the results shown in Fig.~\ref{gastemp} for the same density and \go.
Exposure to \go\ less than 6\ti 10$^4$, as is appropriate to the Ridge, will reduce the gas temperature, as shown 
in this figure.  Thus, we conservatively assume an average gas temperature
of $\,\simlt\,$30$\;$K at an \av\ of 4 and $n$(\mh) between about 3\ti 10$^4\:$\cmc\ and 3\ti 10$^5\:$\cmc.

To assess the impact of gas temperature on the \water\ line flux,
we compute the line intensity within an \av\ bin for a range of temperatures using the one-dimensional non-LTE
radiative transfer code, RADEX \citep{vanderTak07}.  Collisional de-excitation rates for o-\water\ by ortho-\mh\
are those of \citet{Daniel11} and o-\water\ by para-\mh\ are those of \citet{Dubernet09}.  Each point in
Fig.~\ref{ratiofig} panel b corresponds to an \av\ bin of approximately 2.5 magnitudes, or an \mh\ column density
of 2.4\ti 10$^{21}\:$\cms\ per \av\ bin, and an \water\ column density of 2.4\ti 10$^{14}\:$\cms\ per \av\ bin.
These calculations assume 
an \water\ abundance of 10$^{-7}$, though the ratio results presented in Fig.~\ref{exciteration} are not particularly 
sensitive to this assumption.  
Finally, the average \water\ line FWHM within the map area, excluding BN/KL and OMC-2, is $\sim\,$2.2\kms. 
We assume velocity gradients corresponding to line widths of 0.2\kms\ and 1.0\kms\ per \av\ bin.
Fig.~\ref{exciteration} shows the ratio of the \water\ line flux per \av\ bin for a range of gas temperatures 
relative to that at 25$\:$K.  These results suggest that FUV-heated gas at an \av\ of 4 can elevate the
\water\ emission within an \av\ bin by a factor of between $\sim\,$1.2 and 1.4 relative to the \water/\cio\ ratio 
of 0.30\dash 0.35 deep within the cloud.  This suggests that FUV-heated gas alone cannot reproduce the 
\water\ peak in Fig.~\ref{ratiofig}.

\vspace{2mm}

\subsubsection{\it Principal Component Analysis (PCA)}

A second approach to understanding the depth-dependent correlation between species derives from plotting the
\av-ordered integrated intensities of species 1 versus species 2; species whose integrated intensities share a common
trend with increasing \av\ would be highly correlated (as reflected in a high correlation coefficient), while species whose
integrated intensities diverge with depth would have a lower correlation coefficient.  Because the average density of
the Orion Ridge is between $\sim\,$3\ti 10$^4$ and 3\ti 10$^5\:$\cmc,
\ico\ and \cio\ are not considered here due to their 
low critical densities ($n_{\rm crit}\sim\:$2\ti 10$^3\;$\cmc).  All other species have a critical density in excess of
10$^5\;$\cmc\ and are included in this analysis.

Plots of the pairwise correlations between the integrated intensities of the six species considered would require 15 
separate plots.  PCA provides a more compact way to convey
the same information.  Fig.~\ref{pcafig} shows the results for \water, CN, HCN, C$_2$H, N$_2$H$^+$, and NH$_3$.
As explained in greater detail in Paper~I, in PCA, we attempt to explain the total variability of $p$ correlated 
variables through the use of $p$ orthogonal principal components (PC). The components themselves are 
merely weighted linear combinations of the original variables such that PC$\:$1
accounts for the maximum variance in the data of any possible linear combination,
PC$\:$2 accounts for the maximum amount of variance not explained by PC$\:$1
and that it is  orthogonal to PC$\:$1, and so on.  Even though use of all $p$ PC's permits the full
reconstruction of the original data, in many cases, the first few PC's are sufficient to
capture most of the variance in the data.  For the species considered here, PC's 1 and 2 
(Fig.~\ref{pcafig}, left panel) capture 87\% of the total variance and are thus most relevant to our discussion.

There are two key elements in the plot to note: (1) the degree to which
each vector approaches the unit circle and (2) the clustering of vectors.
Because the principal components are normalized such that the quadrature sum of the 
coefficients for each species is unity, the proximity of the points to the circle of unit
radius is a measure of the degree to which any two principal components account for
the total variance in this sample.  Consequently, the closeness of all the points to the
unit circle in the left panel is a reflection of the fact that these two principal 
components contain almost all of the variance in the data, as noted above.

The degree of clustering of the vectors is a measure of their correlation.  Specifically, 
in the limit where two points actually lie on the unit circle, the cosine of the angle
between these points is their linear correlation (see \citeauthor{Neufeld07}~\citeyear{Neufeld07}).
Thus, points that coincide
on the unit circle (i.e., $\Delta\theta =\:$0\ddeg) would indicate perfectly correlated data,
whereas orthogonal vectors (i.e., $\Delta\theta =\:$90\ddeg) would indicate 
perfectly uncorrelated data.  Thus, Fig.~\ref{pcafig} shows that the \water\ distribution is
well correlated with that of CN, HCN, and C$_2$H, which are all predicted to be surface tracers,
and is largely uncorrelated with N$_2$H$^+$ and NH$_3$. 
This confirms the behavior shown in Fig.$\:$\ref{ratiofig} that gas-phase \water\ is mainly found near 
the cloud surface.

We note that a similar study, absent the inclusion of \water\ and NH$_3$, has been conducted toward the Orion B 
molecular cloud by \citet{Pety17}.  The results, which are summarized in Table~5 of their paper, 
are in broad agreement with those presented here.  Specifically, they find that more than 80\% of the
emission from HCN, CN, and C$_2$H arise from gas at \av$\,\leq\,$15, whereas more than 85\% of the line 
emission from N$_2$H$^+$ arises from gas at \av$\,\geq\,$15.

\vspace{2mm}

\subsubsection{{\it H$_2$O Integrated Intensity vs. A$_V$ for a Range of G$_{\rm o}$ and n$_{\rm H}$}}

Grains play a critical role in the computations as they are responsible for extinction of the FUV field
with depth and both provide the surfaces that drive chemistry at moderate cloud depths and serve
as sites for the freezing out of species at large cloud depths.  In particular, the combined effect of
water formation on grains with photodesorption of the water is responsible for the \water\ 
abundance peak at moderate depths seen in the models (see~Hollenbach et al.~Figure 3).  Deep in the cloud, 
the photodesorption rate is negligible, water remains on the grains, and the gas-phase water abundance drops.
However, the grains at large depths in our modeled region are likely above the freeze-out temperature
for CO ($\sim\,$15-20$\:$ K), so CO remains in the gas phase.  Within these regions, He$^+$ attacks CO 
and the resultant atomic O can form water on grain surfaces or, for warmer ($T_{\rm gr}\,\simgt\:$45$\:$K)
grains, in the gas phase, which then freezes onto
grain surfaces, removing O from the gas phase.  This makes the steady-state abundance of gas-phase CO very 
low at large cloud depths, and can even result in a gas-phase elemental carbon abundance that exceeds that of O,
since most of the O is captured in water ice.   However, steady-state models are not appropriate
deep in the cloud, as the He$^+$ arises from cosmic ray ionization of He, and this very
slow rate means that steady state is often not achieved during the lifetime of the cloud.
In fact, our observations show a strong correlation of \ico\ column density with depth, suggesting 
that the clouds we have observed are too young for the abundance of CO to have been diminished
significantly by reactions with He$^+$.

To assess the agreement between models and observations, we compute the o-\water\ 557$\:$GHz integrated
intensity, $\int\!T_{\rm mb}\,dv$, as functions of \av, gas density, and \go\ using the steady-state PDR model described 
in \citet{Hollenbach09}.  This model predicts that gas-phase CO largely disappears in the cloud interior, if not 
from CO freeze-out, then eventually from He$^+$ destruction.  Though He$^+$ destruction of CO leads to a 
gradual increase in the \water-ice abundance deep in the cloud, as described above, the effect on the gas-phase 
\water\ abundance is negligible.  Moreover, because the timescale for \water\ freeze-out, i.e., a few \ti 10$^4$ years, 
is short compared to the age of the cloud, the steady-state model should properly reproduce the gas-phase \water\
abundance.  With these inputs, the model self-consistently computes 
the gas and grain temperatures, and the abundances of more than 60 species in the gas phase and in ice mantles 
on grain surfaces, starting from the cloud surface and extending in to an extinctions of \av$\,=\,$20.
This region is divided into 200 zones of $\delta$\av, each with a fixed density, but with values of gas temperature 
and \water\ column density determined from the model.  The integrated intensities were computed based on the
summed results of RADEX calculations for each zone to a given \av.  

The results are shown in Figs.~\ref{Eb800_Plot} and \ref{Eb1800_Plot} to an \av\ of 10, which accentuates the depths
most sensitive to the FUV field.  In the model of \citet{Hollenbach09}, it was
assumed that the binding energy of atomic oxygen to an interstellar dust grain, $E_{\rm b}$, was 800$\:$K, based on the work of
\citet{Tielens82}.  More recent laboratory work by \citet{He15} determine the binding energy to be closer to 1800$\:$K.
Fig.~\ref{Eb800_Plot} shows the results for the older value of $E_b\,=\,$800$\:$K, while Fig.~\ref{Eb1800_Plot} shows 
the results for the current value of $E_b\,=\,$1800$\:$K.

Several things are clear.  Lower density gas, i.e.,
$n_{\rm H}\:\simlt\:$10$^4$~\cmc, cannot reproduce our results, regardless of the value of \go.  Likewise, 
large values of \go, i.e., \go$\:\simgt\:$10$^4$, offer a poor fit to the data, regardless of
gas density.  The best fit to the majority of data points suggests an average value of \go\ of a few hundred and
a gas density of $\sim\,$3\ti 10$^4$~\cmc, in agreement with previous observations (see \S 3.1.4).

\vspace{2mm}

\subsection{{\it Cepheus B Molecular Cloud}}

As a measure of the gas column density along these strip maps, we used \ico\ $J$\,=\,1\dash 0 data
obtained with FCRAO. These data were obtained using an early 
version of the Sequoia focal plane array receiver when it had only 16 pixels.
The data consisted of an 8\ti 8 point map oriented in RA and Dec with spectra 
spaced by 44.3 arcseconds and the FCRAO telescope at this frequency 
had a FWHM beam size of approximately 47 arcseconds. One of 8 point rows was 
within 2 arcseconds of being aligned
with the northern {\em Herschel} strip map and we plot the integrated intensity of
\ico\ along with that of water and ammonia in Fig.~\ref{stripscannorth}. Unfortunately, the southern 
strip map lies midway between two of the rows of FCRAO data.  Therefore we have
averaged the data from the two rows that lie on either side of the {\em Herschel} strip 
map for comparison. The integrated intensity of the \ico\ line for this average is 
shown in Fig.~\ref{stripscansouth}.

At the declination of the northern strip map, the position of the ionization 
front is located at RA\,(J2000) = 22$^{\rm h}$ 56$^{\rm m}$ 53$^{\rm s}\!\!$.\,7 
based on {\em Spitzer} images (Rob Gutermuth, 
private communication), or at an offset in our strip map of -154 arcseconds.
The northern strip map starts at an offset of -175 arcseconds, and thus well
off any molecular emission. Water emission is detected near the ionization front
and moving to the east; the integrated intensity increases rapidly to a nearly constant 
level with offset: the very strong water emission at offsets greater than +120 
is due to a molecular outflow, which is discussed later. The ammonia emission, on 
the other hand, is not detectable near the ionization front and does not rise to 
a level close to that of water until much further to the east. Based on the
integrated intensity of \ico, the gas column density near the ionization front 
is low and increases steadily moving from west to east.  

The water and ammonia lines arise from rotational levels with nearly the same 
energy above the ground state and both transitions have very similar critical 
densities.  Therefore, these lines have nearly identical excitation conditions.
In the low excitation limit (see Melnick et al. 2011), the conversion factors
between integrated intensity and column density are nearly the same for the
water and ammonia lines.  Thus, the presence of water emission and the absence 
of ammonia emission near the interface must be due to water having a much larger
column density than ammonia.  However, by offsets of +100 arcseconds, both lines have similar
integrated intensity and thus by this position water and ammonia have similar column 
densities.  

The southern strip map (see Fig.~\ref{cephspectrasouth}) unfortunately starts in a position already with
substantial water emission. At the declination of this strip map, the position
of the ionization front is located at RA\,(J2000) = 22$^{\rm h}$ 56$^{\rm m}$ 50$^{\rm s}\!\!$.\,3, 
or at an offset of -233 arcseconds.  The ionization front is approximately 50 arcseconds west
of the start of this strip map, so the strip map does not cover the ionization 
front.  However, as in the northern strip map, the region closest to the ionization
front has strong water emission, small \ico\ gas column density and no 
detectable ammonia emission. Ammonia emission is not detected until much further
east in the strip map where there is much stronger water and CO emission but, again, this is
almost certainly associated with a molecular outflow.

A surprising discovery is the presence of broad water emission toward positions of
several of the spectra in the two strip maps. The most prominent broad-line emission
is seen toward the eastern end of the northern strip map at an offset of +151 arcseconds
(see Fig.~\ref{cephoutflow}).
The water emission in this direction extends over a velocity of 30\kms\
and shows a pronounced self-absorption feature, 
similar to the water spectra seen in many other outflows \citep[see][]{Kristensen11}.
Secondary emission peaks at LSR velocities of $-$45~\kms\ and $+$20~\kms\ are suggestive
of high-velocity \water\ bullets seen in other sources \cite[c.f.][]{Kristensen11a}; however, the
statistical significance of these features, particularly at $+$20~\kms, is low.
The most prominent outflow emission is found toward RA\,(J2000) = 22$^{\rm h}$ 57$^{\rm m}$ 38$^{\rm s}$ 
and Dec\,(J2000) = +62$^{\rm o}$ 35\amin\ 45\asec.  Broad lines are also seen in the spectra further 
east in the northern strip map and in several positions in the southern strip map 
around RA\,(J2000) = 22$^{\rm h}$ 57$^{\rm m}$ 18$^{\rm s}$ and 
Dec\,(J2000) = +62$^{\rm o}$ 34\amin\ 45\asec. The positions where
broad wings are detected suggest the outflow has a angular extent of at least 2.5 arcminutes, 
assuming the broad wing emission is all due to a single molecular outflow.  
It is curious that this region of Cepheus B has been mapped in several transitions 
of \nco\ \citep{Minchin92, Beuther00, Mookerjea06} with no reported mention of 
high-velocity wing emission.

{\em Spitzer} has provided an extensive inventory of the young stars associated with 
this region.  Toward the region of the water outflow is a very prominent source, particularly at the 
longer {\em Spitzer} bands, and is cataloged in the study by \citet{Allen12} and identified as 
a Class I object. This source is associated with Cep OB3 and has a luminosity of about 
200 \lsol\ \citep{Kryukova12}. This source is at RA\,(J2000) = 22$^{\rm h}$ 57$^{\rm m}$ 38.06$^{\rm s}$
and Dec\,(J2000) = +62$^{\rm o}$ 35\amin\ 41.08\asec\ lies toward
the strongest outflow emission and is one of the more luminous sources embedded
in the cloud, with a near-infrared luminosity of about 200 solar luminosities
\citep{Kryukova12}. We suggest that this source is the likely the origin of 
the molecular outflow.

\vspace{2mm}

\section{DISCUSSION}

Water forms in quiescent molecular clouds primarily via two routes.  First, a sequence of gas-phase
ion-neutral reactions beginning with the ionization of \mh\ by cosmic rays or X-rays eventually leads to
the production of H$_3$O$^+$, which is destroyed by dissociative
recombinations yielding 

\vspace{-1.5mm}

\begin{equation}
{\rm H}_3{\rm O}^+ + e^- \rightarrow\ \left\{ \begin{array}{l} {\rm H}_2{\rm O} + {\rm H} \\
{\rm OH} + {\rm H_2}  \\
{\rm OH} + {\rm H} + {\rm H} \\
{\rm O} + {\rm H_2} + {\rm H}~.
\end{array}
\right.
\end{equation}

\vspace{5mm}

With a fractional yield of 0.60$\,\pm\,$0.02, the OH + H + H channel dominates, whereas the branching ratio for water 
production is 0.25$\,\pm\,$0.01 \citep{Jensen00}.  Second, water can form on grain surfaces beginning with gas-phase O atoms 
striking and sticking to grains, followed by a series of surface reactions with H atoms to form OH$_{\rm ice}$ then \water$_{\rm ice}$.

Fig.~\ref{h2oabund} shows the predicted abundance of gas-phase \water\ based on the model of \citet{Hollenbach09}
in which the updated value of the oxygen binding energy to grains of 1800$\:$K is assumed \citep{He15}.  This model
computes the steady-state thermal and chemical structure of a molecular cloud illuminated by an external ultraviolet 
radiation field.  Near cloud surfaces, i.e., \av$\:\simlt\:$1, FUV photons have several effects. 
First, they photodissociate \water\ formed in the gas phase.
Second, they warm the grains, accelerating the thermal desorption of weakly bound O atoms before they can react with
H, thus suppressing the production of \water\ on grain surfaces.  Third, they can photodesorb the ices that do form, placing 
\water\ into the gas phase, which is then subject to photodissociation.  Thus, under a broad range
of gas densities and FUV field strengths, the gas-phase water
abundance near the cloud surface is predicted to be less than 10$^{-8}$ relative to \mh\ \citep[see][]{Hollenbach09}.

Deeper into the cloud, i.e., 1$\:\leq\:$\av$\:\leq\:$10, the FUV field is reduced, but not fully attenuated.  Within this
region, the residual FUV field remains sufficiently strong to photodesorb ices while the photodestruction rate of \water\ 
is reduced.  The combined effect is a peak in the gas-phase \water\ abundance.

Fig.~\ref{h2oabund} shows the profiles of gas-phase water abundance versus depth into the cloud for the current value
of the oxygen binding energy.  These profiles reflect two distinct chemical pathways.
For \go$\:<\:$10$^4$, the peak abundance is set by the balance between O freeze-out, 
hydrogenation, and \water\ photodesorption.  For \go$\:>\:$10$^4$, the grains are sufficiently warm
that O does not freeze out but, instead, is driven into \oxy\ and \water\ in the gas phase. 

These two pathways are understood to result from the OH formation timescale on grain surfaces.  
Fig.~\ref{lifetime_plot} shows the timescale for an H atom to strike a grain, and thus form OH, versus 
grain temperature for an \mh\ density of 3\ti 10$^4$~\cmc, an H-atom density of 1$\:$\cmc, and 
a grain radius of 0.1\um\ \citep[see][]{Hollenbach09}.  Also shown are the timescales for O-atom 
desorption from a grain surface for $E_{\rm b} =\:$800$\:$K and $E_{\rm b} =\:$1800$\:$K.
For a given grain temperature, the higher binding energy allows more time for an H atom to hit the grain 
and relatively quickly form OH before an O atom would otherwise be thermally desorbed.
The main effect is to allow the formation of OH and, ultimately, \water-ice on grains at temperatures as high as
$T_{\rm gr\,}\!\!\sim\,$50$\:$K and \go's as high as 10$^4$ .

Deep into the cloud, i.e., \av$\:\geq\:$\,10, the FUV field is unimportant, the photodesorption rate is negligible,
and \water\ remains frozen out on grain surfaces.  Because the sublimation temperature of \water$_{\rm ice}$ 
exceeds 90$\:$K \citep{Fraser01}, absent an embedded heat source, the water and the oxygen it contains remains trapped on grains and 
unavailable for gas-phase reactions.  
At these large depths, small amounts of gas-phase \water\ (i.e., $\simlt\:$10$^{-9}$ relative to \mh)
are still formed, driven by cosmic-ray-initiated ion-molecule gas-phase reactions.
The profile of \water\ abundance in which the gas-phase water abundance is low at the cloud surface, rises to a 
peak a few \av\ into the cloud, and then drops due to freeze-out is supported by the {\em Herschel} 
data and confirms the results presented in Paper~I, which were based on a much smaller dataset.  

The absence of any abrupt drop in the CO column density deep in the cloud supports data that indicate that
grains remain too warm to allow significant freeze-out, the timescale for which being relatively short compared
to the age of the cloud.  Nevertheless, even if grains are warm enough to keep CO in the gas phase, as noted earlier,
CO destruction can occur via the reaction He$^+$ $\!+\;$CO $\rightarrow$ C$^+~+$ O $+$ He, leading to O going into 
water ice, thus
reducing the atomic oxygen available to reform CO in the gas phase.  As a result, over a long time, the CO
gas-phase abundance drops.  However, cosmic ray rates are slow, especially deep in a cloud, so this process probably 
does not deplete gas-phase CO much during the life of the cloud, which is presumably a few to 10 Myr.

Observational evidence for the above scenario is important for three reasons.  First, as originally noted by 
\citet{Bergin00} and in a large number of subsequent models, the reduced gas-phase O abundance at high 
\av's due to \water\ freeze-out has been invoked to explain the non-detections of \oxy\ toward 
quiescent dense clouds observed by {\em SWAS}, {\em Odin}, and {\em Herschel}.
The large number of lines of sight along which \oxy\ has been searched for and not detected (to abundance levels between
10$^{-7}$ and 10$^{-8}$) testifies to the ubiquity of the above scenario.  Similarly, the few lines of sight toward which
detections of \oxy\ have been reported \citep{Larsson07, Goldsmith11, Liseau12, Chen14} require explanations
other than cold quiescent gas in chemical equilibrium, such as a warm (i.e., \tgas$\;\ge$\, 65 K), dense core or 
shock-excited gas within which the \oxy\ abundance can be enhanced \citep{Melnick15}.

Second, the obervations reported here underscore the danger in deriving gas-phase \water\ abundances based on the
ratio of \water\ to CO (or \ico\ or \cio) column densities.  Because gas-phase \water\ is more of a surface tracer than CO, 
it is likely that gas-phase water abundances derived in this manner will underestimate the peak water abundance
in regions where \water\ is largely in the gas phase, 
particularly toward clouds possessing large CO column densities (i.e., corresponding to \av$\,\simgt\:$10).

Third, the presence of gas-phase \water\ predominantly near cloud surfaces, as confirmed by the data presented here,
indirectly supports the prediction that water ice is quite abundant at depths greater than $\sim\,$5\dash 10 \av\ magnitudes
into dense clouds.  Limited observations by {\em ISO}, {\em Spitzer}, and {\em Akari}, i.e., less than approximately
250 lines of sight in total, support this conclusion through direct measures of near- and mid-infrared water-ice absorption
features (see \citealt{Gibb04}; \citealt[][and references therein;]{Boogert15} \citealt{Aikawa12, Noble13}).  This
prediction will be subject to further test by future NASA missions, such as {\em SPHEREx} and the 
{\em James Webb Space Telescope} ({\em JWST}).
In particular, {\em SPHEREx} will obtain 0.75 \dash 5.0\um\ absorption spectra toward more than 20,000 
(and as many as 2\,$\times$\,10$^6$) lines of sight within the Milky Way possessing strong indications of intervening 
gas and dust toward spatially isolated background stars, including sources at a variety of evolutionary stages (e.g., diffuse clouds,
dense clouds, young stellar objects, and protoplanetary disks). 
At the same time, in addition to its ability to survey relatively small areas for ice absorption, {\em JWST} 
will be able to follow up select {\em SPHEREx}-identified sources with higher spectral resolving power, 
measure weak ice absorption features with greater sensitivity, and, complement the $\lambda \leq\;$5\um\ ice measurements 
of {\em SPHEREx} with measures of the ice features beyond 5\um.  With the added data from these missions, 
we should have the clearest picture yet of how water is distributed in interstellar clouds.

\vspace{20mm}

%\section{SUMMARY}

Support for this work was provided by NASA Astrophysics Data Analysis Program (ADAP) 
grant NNX13AF16G and an award issued by JPL/Caltech.  Part of this
research was carried out in part at the Jet Propulsion Laboratory, which is operated for NASA
by the California Institute of Technology.
J.R.G. thanks the Spanish MICIU for funding support under grant AYA2017-85111-P.

%grant 1329220.

\clearpage

\bibliographystyle{plainnat}

\begin{thebibliography}{}

\vspace{1.7mm}

\bibitem[Aikawa \etal(2012)]{Aikawa12} Aikawa, Y., Kamuro, D., Sakon, I., \etal\ 2012, \aap, 538, A57

\bibitem[Allen \etal(2012)]{Allen12} Allen, T.S., Gutermuth, R.A., Kryukova, E., \etal\ 2012, \apj, 750, 125

\bibitem[Bally \etal(1987)]{Bally87} Bally, J., Langer, W. D., Stark, A. A., \etal\ 1987, \apj, 312, L45

\bibitem[Bergin \etal(1994)]{Bergin94} Bergin, E. A., Goldsmith, P. F., Snell, R. L., \etal\ 1994, \apj, 431,
674

\bibitem[Bergin \etal(2000)]{Bergin00} Bergin, E. A., Melnick, G. J., Stauffer, J.R., \etal\ 2000, \apj, 539, L132

\bibitem[Bergin \etal(1996)]{Bergin96} Bergin, E. A., Snell, R. L., \& Goldsmith, P. F. 1996, \apj, 460, 343

\bibitem[Beuther \etal(2000)]{Beuther00} Beuther, H., Kramer, C., Deiss, B., \etal\ 2000, \aap, 362, 1109

\bibitem[Bisbas \etal(2012)]{Bisbas12} Bisbas, T.G., Bell, T.A., Viti, S., \etal\ 2012, \mnras, 427, 2100

\bibitem[Boogert, Gerakines, \& Whittet(2015)]{Boogert15} Boogert, A.C.A., Gerakines, P.A., \& Whittet, D.C.B. 2015,
\araa, 53, 541

\bibitem[Caselli \etal(2012)]{Caselli12} Caselli, P., Keto, E., Bergin, E.A., \etal\ 2012, \apj, 759, L37

\bibitem[Cernicharo \etal(2006)]{Cernicharo06} Cernicharo, J.,  Goicoechea, J.R., Pardo, J.R., \etal\ 2006, \apj, 642, 940

\bibitem[Chen \etal(2014)]{Chen14} Chen, J.-H., Goldsmith, P.F., Viti, S., \etal\ 2014, \apj, 793, 111

\bibitem[Coutens \etal(2012)]{Coutens12} Coutens, A., Vastel, C., Caux, E., \etal\ 2012, \aap, 539, A132

\bibitem[Danby \etal(1988)]{Danby88} Danby, G., Flower, D.R., Valiron, P., \etal\ 1988, \mnras, 235, 229

\bibitem[Daniel, Dubernet, \& Grosjean(2011)]{Daniel11} Daniel, F.,  Dubernet, M., \& Grosjean, A. 2011, 
\aap, 536, A76

\bibitem[Daniel \etal(2005)]{Daniel05} Daniel, F., Dubernet, M.-L., Meuwly, M., \etal\ 2005, \mnras, 363, 1083

\bibitem[Dubernet \etal(2009)]{Dubernet09} Dubernet, M.-L., Daniel, F., Grosjean, A. \etal\ 2009, \aap, 497, 911

\bibitem[Dumouchel, Faure, \& Lique(2010)]{Dumouchel10} Dumouchel, F., Faure, A., \& Lique, F. 2010, 
\mnras, 406, 2488

\bibitem[Dutrey \etal(1991)]{Dutrey91} Dutrey, A., Langer, W. D., Bally, J., \etal\ 1991, \aap, 247, L9

\bibitem[Fraser \etal(2001)]{Fraser01} Fraser, H.J., Collings, M.P., McCoustra, M.R.S., \etal\ 2001, \mnras, 327, 1165

\bibitem[Gibb \etal(2004)]{Gibb04} Gibb, E.L., Whittet, D.C.B., Boogert, A.C.A., \etal\ 2004, \apjs, 151, 35

\bibitem[Goldsmith \etal(2011)]{Goldsmith11} Goldsmith, P.F., Liseau, R., Bell, T., \etal\ 2011, \apj, 737, 96

\bibitem[Habing(1968)]{Habing68} Habing, H. J. 1968, {\em Bull. Astron. Inst. Netherlands}, 19, 421

\bibitem[Hawkins \& Jura(1987)]{Hawkins87} Hawkins, I., \& Jura, M. 1987, \apj, 317, 926

\bibitem[He \etal(2015)]{He15} He, J., Shi, J., Hopkins, T., \etal\ 2015, \apj, 801, 120

\bibitem[Hollenbach \etal(2009)]{Hollenbach09} Hollenbach, D.J., Kaufman, M. J., Bergin, E.A., \etal\ 2009, \apj, 690, 1497

\bibitem[Jensen \etal(2000)]{Jensen00} Jensen, M.J., Bilodeau, R.C., Safvan, C.P., \etal\ 2000, \apj, 543, 764

\bibitem[Johnstone \& Bally(1999)]{Johnstone99} Johnstone, D., \& Bally, J. 1999, \apj, 510, L49

\bibitem[Kainulainen, Lehtinen, \& Harju(2006)]{Kainulainen06} Kainulainen, J.,
Lehtinen, K. \& Harju, J. 2006, \aap, 447, 597

\bibitem[Kester, Higgins \& Teyssier(2017)]{Kester17} Kester, D., Higgins, R. \& Teyssier, D. 2017, \aap, 599, 115 

\bibitem[Keto, Rawlings \& Caselli(2014)]{Keto14} Keto, E., Rawlings, J. \& Caselli, P.  2014, \mnras, 440, 2616

\bibitem[Keto \& Rybicki(2010)]{Keto10} Keto, E., \& Rybicki, G. 2010, \apj, 716, 1315

\bibitem[Kristensen \& van Dishoeck(2011)]{Kristensen11} Kristensen, L. E. \& van Dishoeck 2011, Astron. Nachr., 332, No.$\:$5, 475

\bibitem[Kristensen \etal(2011)]{Kristensen11a} Kristensen, L. E, van Dishoeck, Tafalla, M., \etal\ 2011, \aap, 531, L1

\bibitem[Kryukova \etal(2012)]{Kryukova12} Kryukova, E., Megeath, S. T., Gutermuth, R. A.,  \etal\ 2012, \aj, 144, 31

\bibitem[Langer \& Penzias(1990)]{Langer90} Langer,W. D., \& Penzias, A. A. 1990, \apj, 357, 477

\bibitem[Larsson \etal(2007)]{Larsson07} Larsson, B., Liseau, R., Pagani, L., \etal\ 2007, \aap, 466, 999

\bibitem[Lee \etal(2014)]{Lee14} Lee, S., Lee, J.-E., Bergin, E.A., \etal\ 2014, \apjs, 213, 33

%\bibitem[Linke \etal(1977)]{Linke77} Linke, R.A, Goldsmith, P.F., Wannier, P.G., \etal\ 1977, \apj, 214, 50

\bibitem[Lique \etal(2010)]{Lique10} Lique, F., Spielfiedel, A., Feautrier, N., \etal\ 2010, {\em J.~Chem.~Phys.}, 132, 024303

\bibitem[Liseau \etal(2012)]{Liseau12} Liseau, R., Goldsmith, P.F., Larsson, B., \etal\ 2012, \aap, 541, 73

\bibitem[Loughnane \etal(2012)]{Loughnane12} Loughnane, R. M., Redman, M. P., Thompson, M. A., \etal\ 2012, 
\mnras, 420, 1367

\bibitem[Melnick \& Kaufman(2015)]{Melnick15} Melnick, G.J., \& Kaufman, M.J. 2015, \apj, 806, 227

\bibitem[Melnick \etal(2011)]{Melnick11} Melnick, G. J., Tolls, V., Snell, R. L., \etal\ 2011, \apj, 727, 13

\bibitem[Minchin, Ward-Thompson, \& White(1992)]{Minchin92} Minchin, N. R., Ward-Thompson, D., \& White, G. J. 1992, \aap, 265, 733

\bibitem[Mookerjea \etal(2006)]{Mookerjea06} Mookerjea, B., Kramer, C., R\"{o}llig, M., \etal\ 2006, \aap, 456, 235

\bibitem[Mottram \etal(2013)]{Mottram13} Mottram, J.C., van Dishoeck, E.F., Schmalz, M., \etal\ 2013, \aap, 558, A126

\bibitem[Mueller \etal(2014)]{Mueller14} Mueller, M., Jellema, W., Olberg, M. \etal\ ``The HIFI Beam: 
Release \#1 Release Note for Astronomers," ESA Doc.: HIFI-ICC-RP-2014-001, v1.1, 1 Oct 2014

\bibitem[Mullins \etal(2016)]{Mullins16} Mullins, A.M., Loughnane, R.M., Redman, M.P., \etal\ 2016, \mnras, 459, 2882

\bibitem[Neufeld \etal(2007)]{Neufeld07} Neufeld, D. A., Hollenbach, D. J., Kaufman, M. J., \etal\ 2007, \apj, 664, 890

\bibitem[Noble \etal(2013)]{Noble13} Noble, J. A., Fraser, H. J., Aikawa, Y., \etal\ 2013, \apj, 775, 85

\bibitem[Ott(2010)]{Ott10} Ott, S.  2010, {\em Astronomical Data Analysis Software and Systems XIX}, 434, 139

\bibitem[Pabst \etal(2019)]{Pabst19} Pabst, C. H. M., Goicoechea, D., Teyssier, \etal\ 2019, {\em in preparation}

\bibitem[Pety \etal(2017)]{Pety17} Pety, J., Guzm\'{a}n, V. V., Orkisz, J. H., \etal\ 2017, \aap, 599, 98

%\bibitem[Pilbratt \etal(2010)]{Pilbratt10} Pilbratt, G. L., \etal\ 2010, \aap, 518, L1

\bibitem[Pirogov(1999)]{Pirogov99} Pirogov, L., 1999, \aap, 348, 600

\bibitem[Pirogov \etal(2003)]{Pirogov03} Pirogov, L., Zinchenko, I., Caselli, P., \etal\ 2003, \aap, 405, 639

\bibitem[Reid \etal(2009)]{Reid09} Reid, M.J., Menten, K.M., Zheng, X.W., \etal\ 2009, \apj, 700, 137

\bibitem[Ripple \etal(2013)]{Ripple13} Ripple, F., Heyer, M.H., Gutermuth, R., \etal\ 2013, \mnras, 431, 1296

\bibitem[Roelfsema \etal(2012)]{Roelfsema12} Roelfsema, P. R., Helmich, F. P., Teyssier, D., et al. 2012, 
\aap, 537, A17 

\bibitem[R\"{o}llig \etal(2007)]{Rollig07} R\"{o}llig, M., Abel, N. P., Bell, T. \etal\  2007, \aap, 467, 187

\bibitem[Savage \etal(2002)]{Savage02} Savage, C., Apponi, A.J., Ziurys, L.M., \etal\ 2002, \apj, 578, 211

\bibitem[Schilke \etal(1992)]{Schilke92}Schilke, P., Walmsley, C.M., Pineau Des For\^{e}ts, G., \etal\ 1992, \aap, 256, 595

\bibitem[Schmalzl \etal(2014)]{Schmalzl14} Schmalzl, M., Visser, R., Walsh, C., \etal\ 2014, \aap, 572, A81

\bibitem[Snell \etal(2000)]{Snell00} Snell, R. L., Howe, J. E., Ashby, M. L. N., \etal\ 2000, \apj, 539, L93

\bibitem[Spielfiedel \etal(2012)]{Spielfiedel12} Spielfiedel A., Feautrier N., Najar F., \etal\ 2012, \mnras, 421, 1891

\bibitem[Stacey \etal(1993)]{Stacey93} Stacey, G.J., Jaffe, D.T., Geis, N., \etal\ 1993, \apj, 404, 219

\bibitem[Tatematsu \etal(1993)]{Tatematsu93}Tatematsu, K., Umemoto, T., Murata, Y., \etal\ 1993, \apj, 404, 643

\bibitem[Tielens \& Hagen(1982)]{Tielens82} Tielens, A. A., \& Hagen, W. 1982, \aap, 114, 245

\bibitem[Ungerechts \etal(1997)]{Ungerechts97} Ungerechts, H., Bergin, E. A., Goldsmith, P. F., \etal\ 1997, \apj, 482, 245

\bibitem[van der Tak \etal(2007)]{vanderTak07} van der Tak, F. F. S., Black, J.H., Sch\"{o}ier, F.L. \etal\ 2007, \aap, 468, 627

\bibitem[Womack \etal(1992)]{Womack92} Womack, M., Ziurys, L.M., Wyckoff, S., \etal\ 1992, \apj, 387, 417

\bibitem[Yang \etal(2010)]{Yang10} Yang, B., Stancil, P.C., Balakrishnan, N., \etal\ 2010, \apj, 718, 1062


\end{thebibliography}

\clearpage

\begin{sidewaystable}[ht]
\begin{center}
{\bf Table 1} \\*[1.9mm]
Spectral Lines Observed by {\sl Herschel/HIFI} and FCRAO \\*[1.3mm]
\begin{tabular}{lcccccc} \hline \\*[-4.3mm]  \hline
\multicolumn{6}{c}{\rule{0mm}{5.5mm}\sl Herschel\,/\,HIFI} \\*[1.2mm] \hline
\rule{0mm}{6mm}  &  &  Energy Above  &  Rest  &  Critical  &  FWHM  & Main Beam\\
Species  &  Transition  &  Ground State  &  ~~Frequency~~  & ~~~Density$^{\,b}$~  &  Beamsize & ~Efficiency$^{\,c}$ \\
  &  &  ($E_u/k$)  &  (GHz)  &  (cm$^{-3}$)  &  (arcsec)  &  ($\eta_{\rm mb}$)\\*[1.2mm] \hline
\rule{0mm}{6.2mm}~H$_2^{\;16}$O  & 1$_{10}$\dash 1$_{01}$ &  27~K  &  556.936  &  1\ti 10$^8$  &  39  & 0.62\\*[1.8mm]
~NH$_3$  & $J,K =\;$1,0\dash 0,0  &  27~K  &  572.498  &  3\ti 10$^7$  &  38  & 0.62 \\*[1.8mm] \hline
\multicolumn{6}{c}{\rule{0mm}{5.5mm}FCRAO} \\*[1.2mm] \hline
\rule{0mm}{6.2mm}~C$_2$H  &  $N =\;$1\dash 0, $J =\,\frac{1}{2}$\dash $\frac{1}{2}$  &  ~~4.2~K  &  
~~\hspace{2.0mm}87.402$^{\,a}$  &  2\ti 10$^5$  &  56 & 0.48 \\*[1.8mm]
~HCN  &  $J =\:$1\dash 0  &  ~~4.3~K  &  ~~\hspace{2.0mm}88.632$^{\,a}$  &  3\ti 10$^6$  &  55 & 0.48 \\*[1.8mm]
~N$_2$H$^+$  &  $J =\:$1\dash 0  &  ~~4.5~K  &  ~~\hspace{2.0mm}93.174$^{\,a}$  &  4\ti 10$^5$  &  52 & 0.47 \\*[1.8mm]
~C$^{18}$O  &  $J =\:$1\dash 0  &  ~~5.3~K  &  109.782  &  2\ti 10$^3$  &  46 & 0.45 \\*[1.8mm]
~$^{13}$CO  &  $J =\:$1\dash 0  &  ~~5.3~K  &  110.201  &  2\ti 10$^3$  &  46 & 0.48 \\*[1.8mm]
~CN  &  $N =\;$1\dash 0, $J =\,\frac{3}{2}$\dash $\frac{1}{2}$   &  ~~5.5~K  &  \hspace{2.0mm}113.491$^{\,a}$  &  
2\ti 10$^6$  &  46 & 0.43 \\*[1.8mm] 
~$^{12}$CO  & $J =\:$1\dash 0  &  ~~5.3~K  &  115.271  &  2\ti 10$^3$  &  46 & 0.43 \\*[1.8mm] \hline
\end{tabular}
\end{center}
\vspace{-6mm}
\begin{list}{}{\leftmargin 0.85in \rightmargin 0.70in \itemindent -0.13in}
\item {$^{\rm a}$}~Rest frequency of the strongest hyperfine component.\\*[-7.5mm]
\item {$^{\rm b}$}~Based on the published or interpolated collisional de-excitation rates 
at 30$\,$K and assuming collisions with ortho- and para-\mh\ in the ratio of 0.03, the LTE
value at 30$\:$K.  The rates used are: \water, \citet{Daniel11}; NH$_3$, \citet{Danby88}; 
C$_2$H, \citet{Spielfiedel12};
HCN, \citet{Dumouchel10}; N$_2$H$^+$, \citet{Daniel05}; \ico\ and \cio, \citet{Yang10}; CN,
\citet{Lique10}.\\*[-7.5mm]
\item {$^{\rm c}$}~The HIFI main beam efficiencies are obtained from the calibration data (HIFI\_CAL\_25\_0) of the 
HIFI observation context, ObsID: 1342228626 (SPG Version v14.1.0), which are based on the values published in 
"The HIFI Beam:~Release$\:$\#1 Release Note for Astronomers," Esa Doc.: HIFI-ICC-RP-2014-001, v1.1, 1 Oct 2014.  \\*[-5mm]
\end{list}
\label{lines}
\end{sidewaystable}

\clearpage

\setcounter{figure}{0}

\begin{figure}[ht]
\centering
\vspace{2.5mm}
\hspace{-1.5mm}\includegraphics[scale=0.905]{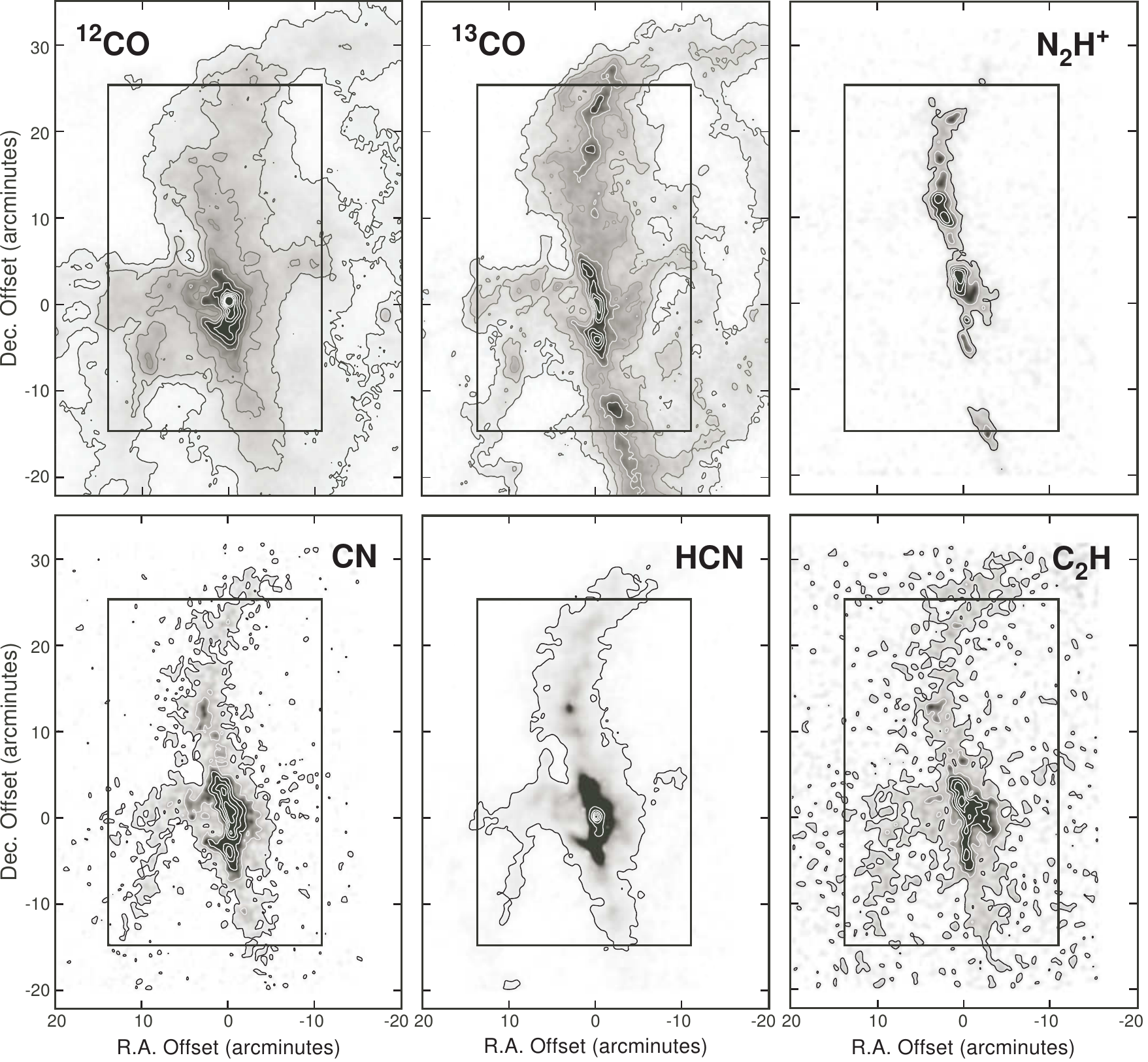}
\vspace{2.7mm}
\caption{Maps of the Orion Molecular Ridge obtained with 46\asec\dash$\:$60\asec\ spatial resolution using the 
FCRAO (see Table~1).  The peak integrated 
intensities ($\int\!T_{\rm mb}\:dv$), in K km s$^{-1}$ are: 700.00 (\nco),
89.41 (\ico), 30.13 (N$_2$H$^+$), 69.42 (CN), 677.8 (HCN),
13.66 (C$_2$H). Contours superposed on the \ico\ and \cio\ maps are in units of 0.10 
of the peak value, with the peak contour shown being 0.9.  
Contours superposed on the N$_2$H$^+$, CN, HCN, and C$_2$H maps 
are in units of 0.15 of the peak value.  The rectangular region outlined in each map
encompasses the region mapped in the \nwater\ 1$_{10}$-1$_{01}$ 556.9~GHz
transition by \hhifi.  
All map offsets are relative to $\alpha\,=\,$05$^{\rm h}$35$^{\rm m}$14$^{\rm s}\!\!.$5,
$\delta\:=\,-$05\ddeg 22\amin 37\asec\ (J2000). }
\label{fcraomaps}
\end{figure}

\clearpage

\begin{sidewaysfigure}[ht]
\centering
\includegraphics[scale=0.835]{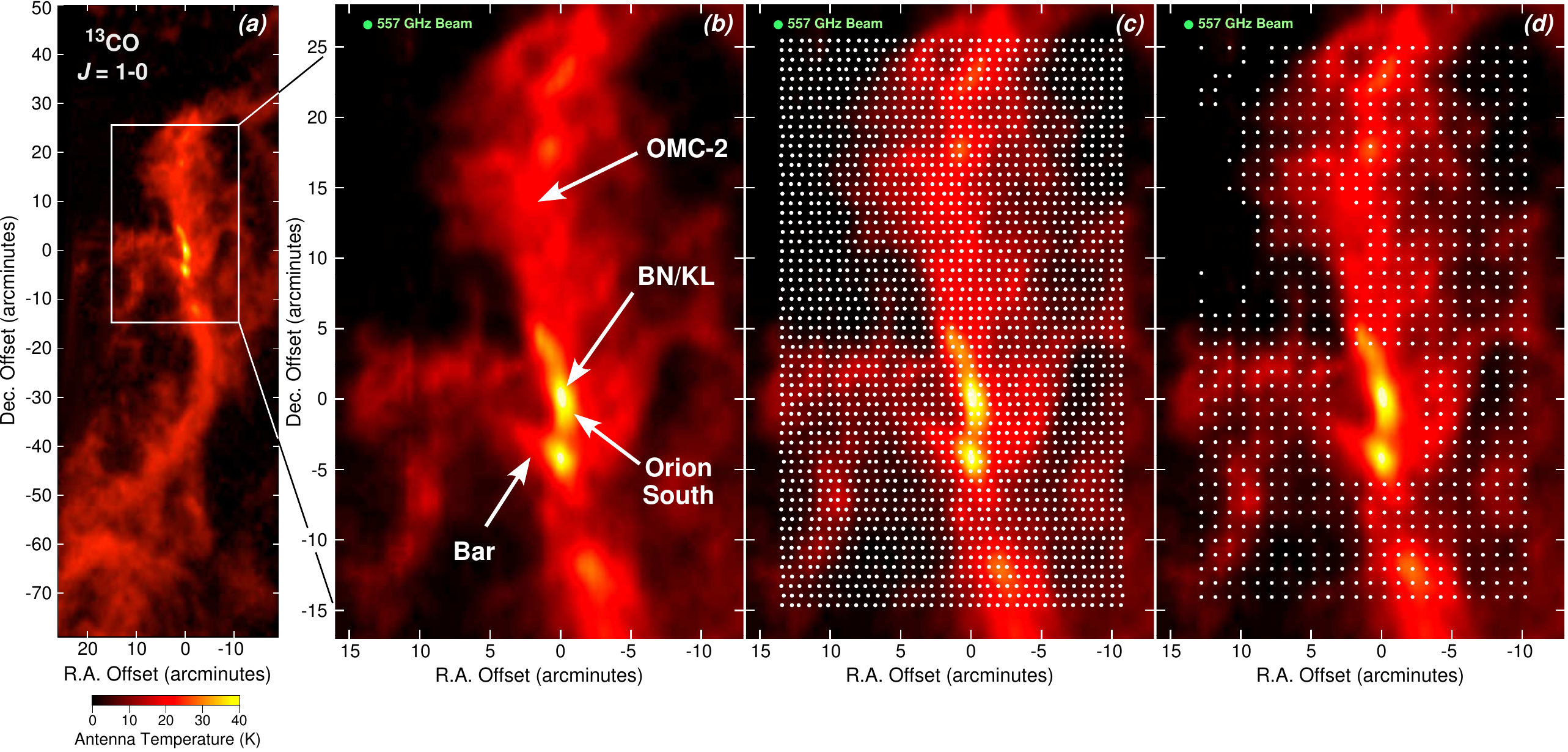}
\vspace{-2.7mm}
\caption{Region mapped by \hhifi\ in this study.  Panel {\em (a)}: \ico\ $J\!=\,$1\dash 0 map of the
Orion complex obtained with the Five College Radio Astronomical Observatory.  The portion of Orion
mapped here is enclosed within the white box.  Panel {\em (b)}: expanded view of the mapped region
with important subregions denoted for reference.  The size of the HIFI beam at 557~GHz is shown
in green at the top.  Panel {\em (c)}: expanded view with the as-observed
positions of the \water\ \tro\ 556.936~GHz and NH$_3$ $J,K =\;$1,0\dash 0,0 572.498~GHz maps
denoted with white dots.  Panel {\em (d)}: regridded positions on which all data
were placed (see text).  A total of 960 positions, separated by 1\amin, were thus obtained; however, only data
toward 834 positions are considered here, since some positions exhibited no detectable \ico\ (or other)
line emission, or because some lines of sight are dominated by outflow emission (e.g., near BN/KL and OMC-2)
rather than quiescent gas emission.  The
regridded positions included in this study are denoted by white dots.  All map offsets are relative to 
$\alpha\,=\,$05$^{\rm h}$35$^{\rm m}$14$^{\rm s}\!\!.$5,
$\delta\:=\,-$05\ddeg 22\amin 37\asec\ (J2000). }
\label{regridpos}
\end{sidewaysfigure}

\clearpage

\begin{figure}[ht]
\centering
\vspace{0.55in}
$\!\!\!$\includegraphics[scale=0.89]{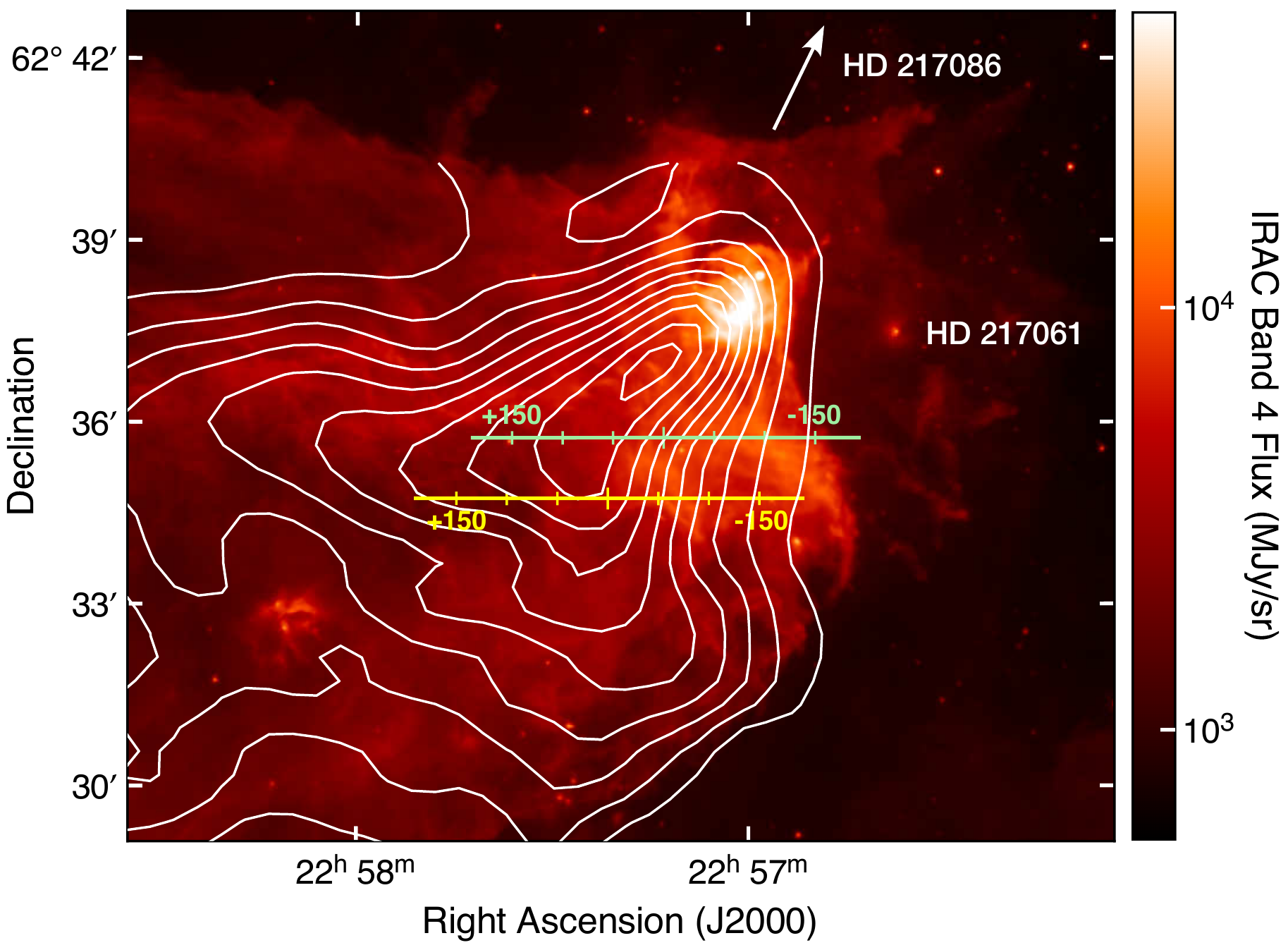}
\vspace{1.0mm}
\caption{{\em Spitzer}/IRAC Band 4 (7.87\um) image of Cepheus B with the contours of \nco\ J\,=\,3\dash 2 emission 
obtained by \citet{Beuther00} superposed. The northern and southern strip scans obtained by {\em Herschel}/HIFI 
in the 557~GHz H$_2^{\:16}$O 1$_{10}$\dash 1$_{01}$ and the 572~GHz NH$_3$ J,K$\,=\,$1,0\dash 0,0 lines 
are also shown with offsets from the scan centers (see text) noted in arcseconds.  The exciting stars, HD\,217061
and HD\,217086 (beyond the top edge of the image), are noted.}
\label{cephscanfig}
\end{figure}

\clearpage

\begin{figure}[ht]
\centering
\vspace{-0.25in}
%$\!\!$\includegraphics[angle=90,scale=0.88]{Cepheus_Water_Spectra_North.pdf}
$\!\!\!\!$\includegraphics[scale=0.60]{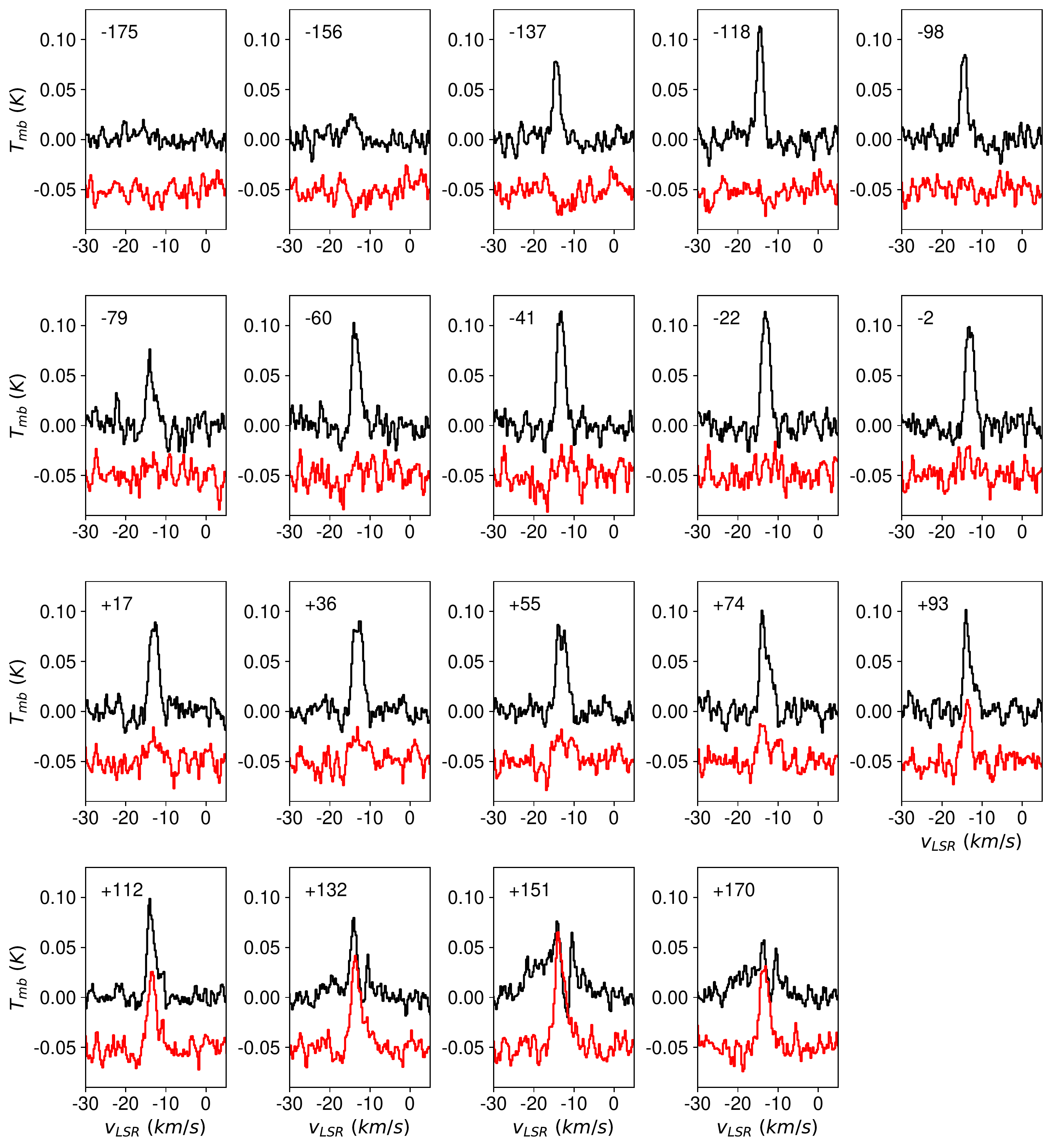}
\vspace{-2.0mm}
\caption{Spectra of \water\ (black) and NH$_3$ (red) along the northern Cepheus B strip scan.  The offsets, in arcseconds, relative to the
(0,0) position of R.A. $=$ 22$^{\rm h}$ 57$^{\rm m}$ 16$^{\rm s}$ and Dec. $= +$62\ddeg\ 35\amin\ 45\asec\ (J2000) are given
in the upper left of each panel.}
\label{cephspectranorth}
\end{figure}

\clearpage

\begin{figure}[ht]
\centering
\vspace{-0.25in}
%$\!\!$\includegraphics[angle=90,scale=0.88]{Cepheus_Water_Spectra_South.pdf}
$\!\!\!\!$\includegraphics[scale=0.60]{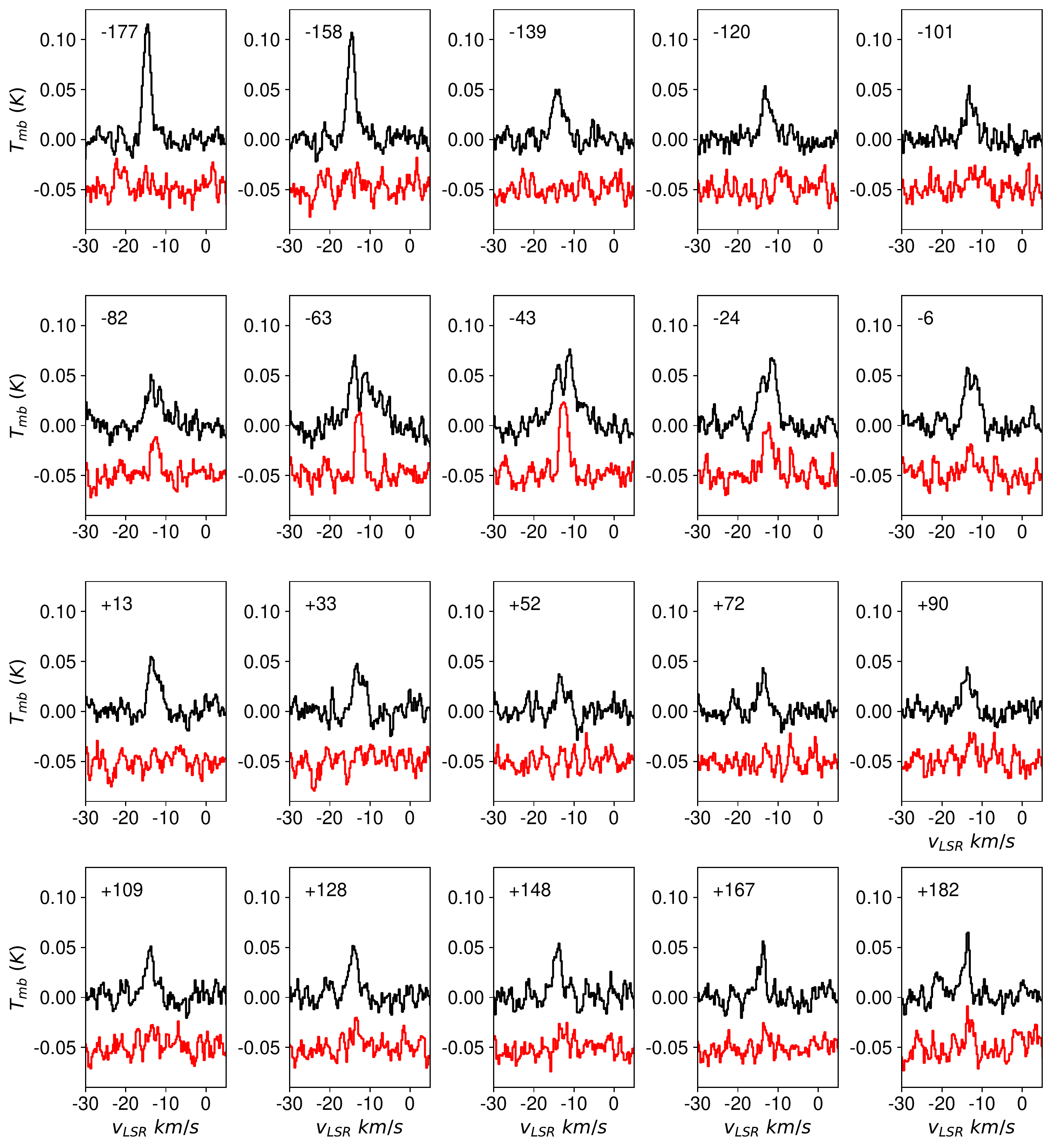}
\vspace{-2.0mm}
\caption{Spectra of \water\ (black) and NH$_3$ (red) along the southern Cepheus B strip scan.  The offsets, in arcseconds, relative to the
(0,0) position of R.A. $=$ 22$^{\rm h}$ 57$^{\rm m}$ 24$^{\rm s}$ and Dec. $= +$62\ddeg\ 34\amin\ 45\asec\ (J2000) are given
in the upper left of each panel.}
\label{cephspectrasouth}
\end{figure}

\clearpage

\begin{figure}[ht]
\centering
\vspace{0.95in}
%$\!\!$\includegraphics[angle=90,scale=0.88]{Cepheus_Water_Spectra_South.pdf}
$\!\!$\includegraphics[scale=0.98]{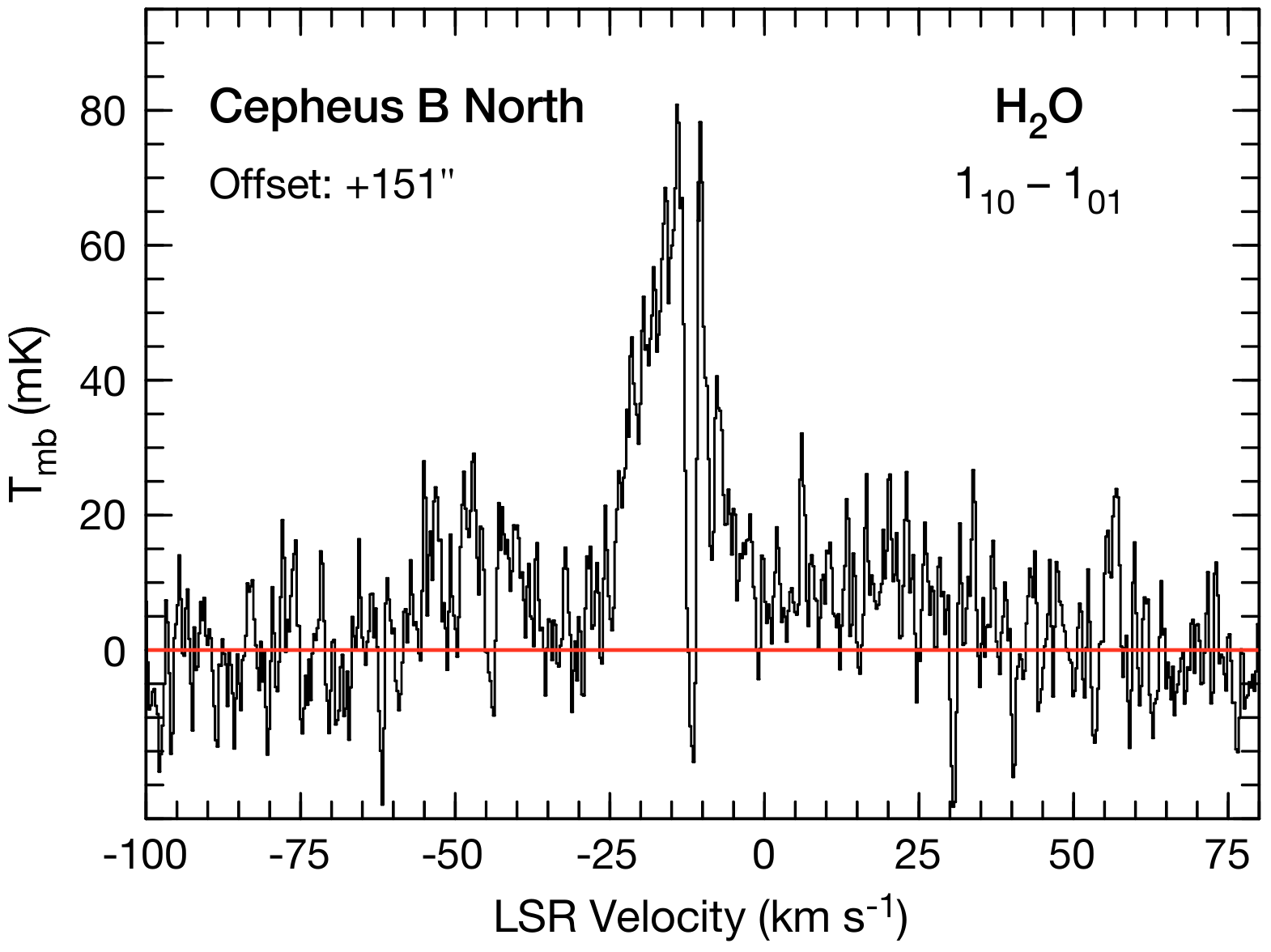}
\vspace{5.0mm}
\caption{Blowup of the continuum-subtracted \water\ spectrum along the Cepheus B northern strip scan
showing evidence of an embedded outflow
at RA $=$ 22$^{\rm h}$ 57$^{\rm m}$ 38$^{\rm s}$ 
and Dec. $= +$62$^{\rm o}$ 35\amin\ 45\asec\ (J2000), corresponding to an offset of $+$\,151\asec\ relative 
to the (0,0) position of R.A. $=$ 22$^{\rm h}$ 57$^{\rm m}$ 24$^{\rm s}$ and Dec. $= +$62\ddeg\ 34\amin\ 45\asec\ 
(J2000).  The continuum-subtracted baseline is shown in red.
}
\label{cephoutflow}
\end{figure}

\clearpage

\begin{figure}[ht]
\centering
\vspace{-2mm}
\includegraphics[scale=0.95]{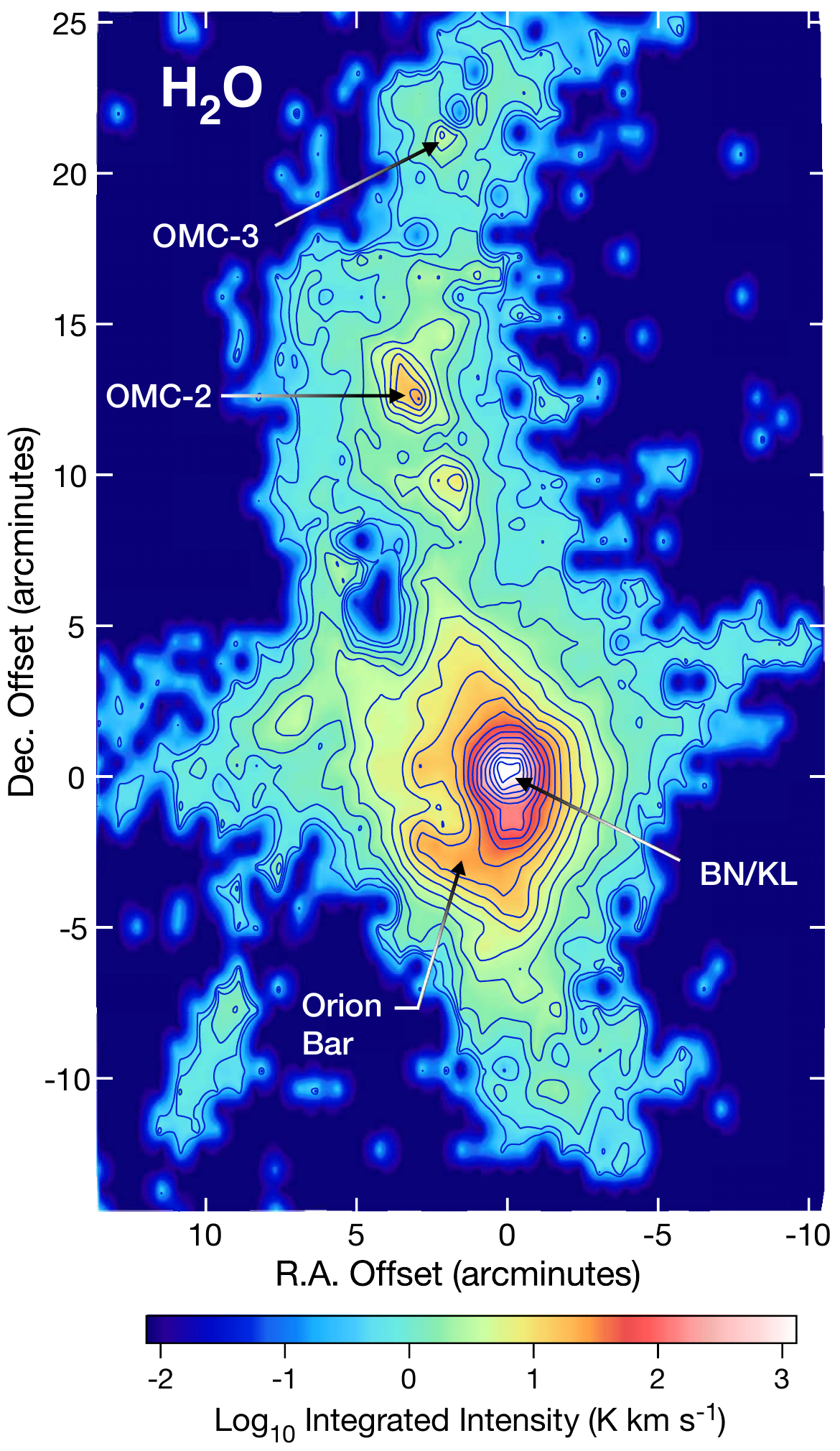}
\vspace{2.2mm}
\caption{\hhifi\ 556.936 GHz \water\ \tro\ integrated intensity map toward Orion.  Note that, because of 
the large range of integrated intensities between the BN/KL region and the extended molecular ridge, the 
map is represented using a logarithmic scale.  All map offsets are relative to 
$\alpha\,=\,$05$^{\rm h}$35$^{\rm m}$14$^{\rm s}\!\!.$5,
$\delta\:=\,-$05\ddeg 22\amin 37\asec\ (J2000). }
\label{h2omap}
\end{figure}

\clearpage

\begin{figure}[ht]
\centering
\vspace{-2mm}
\includegraphics[scale=0.95]{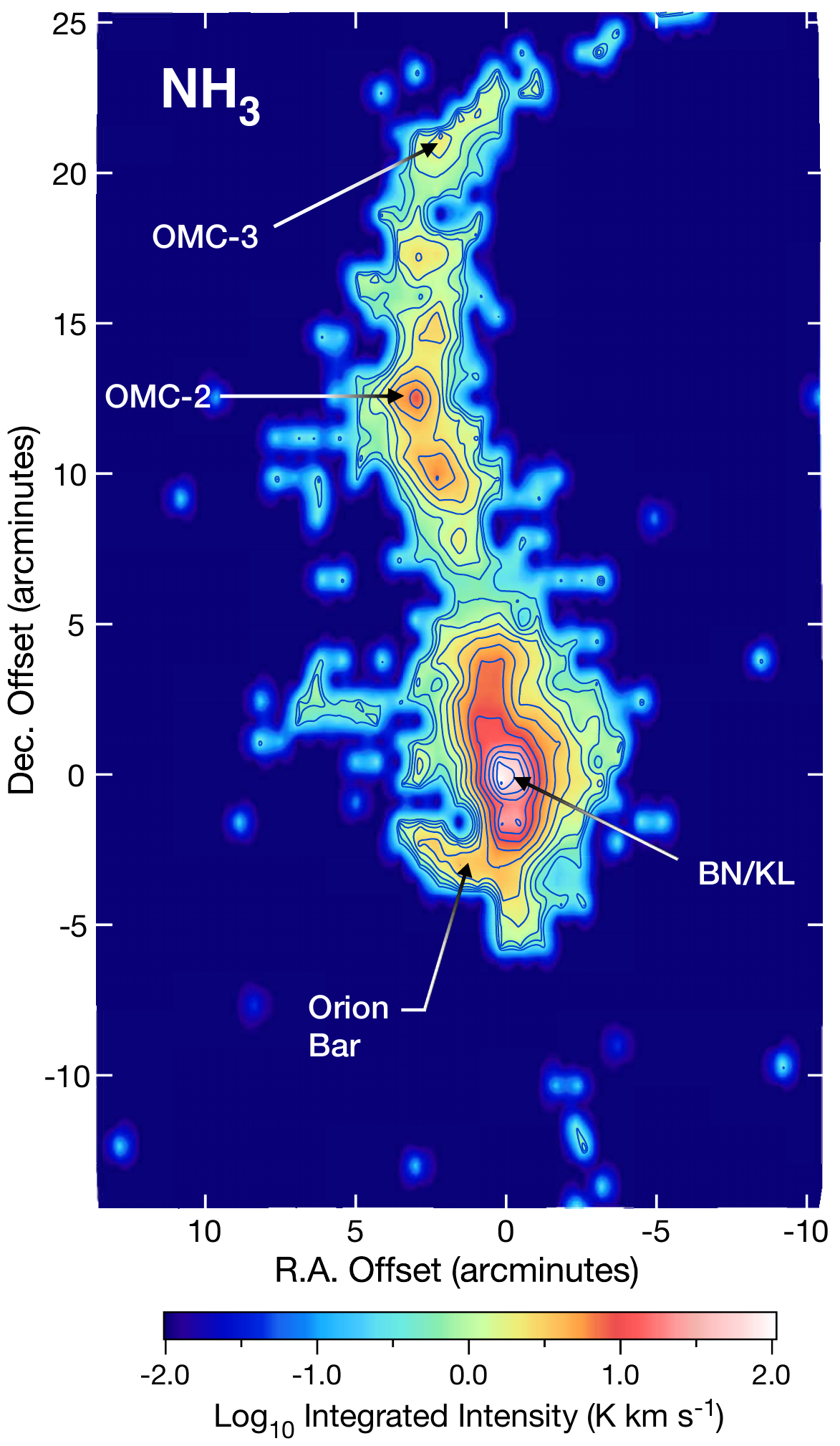}
\vspace{2.2mm}
\caption{\hhifi\  572.498 GHz NH$_3$ $J$\,$=\,$1\,$\rightarrow$\,0 integrated intensity map toward Orion.  
Note that, because of the large range of integrated intensities between the BN/KL region and the extended 
molecular ridge, the map is represented using a logarithmic scale.  All map offsets are relative to 
$\alpha\,=\,$05$^{\rm h}$35$^{\rm m}$14$^{\rm s}\!\!.$5,
$\delta\:=\,-$05\ddeg 22\amin 37\asec\ (J2000). }
\label{nh3map}
\end{figure}

\clearpage

\begin{figure}[ht]
\centering
\vspace{-1mm}
\includegraphics[scale=0.81]{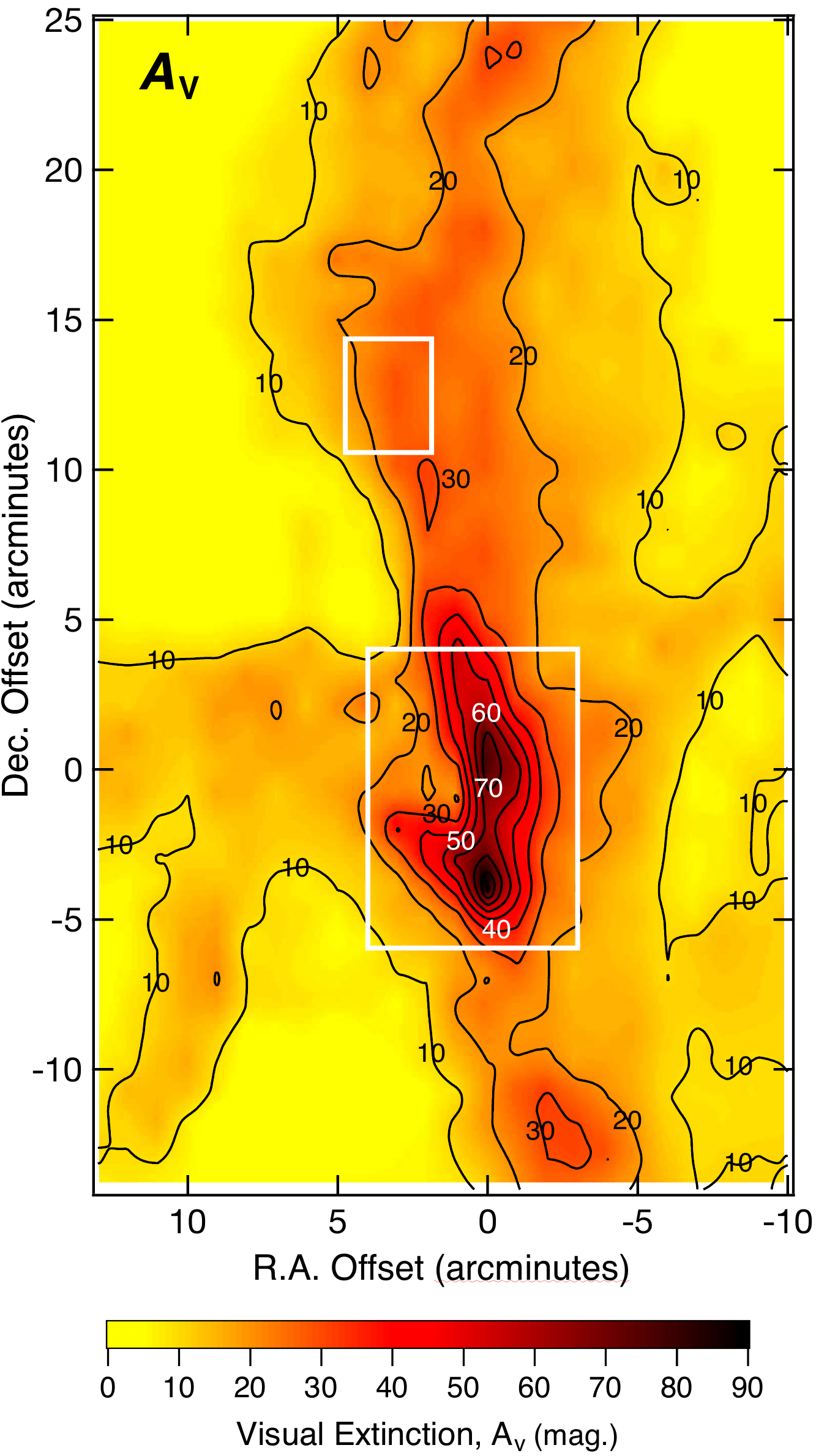}
\vspace{3.6mm}
\caption{Spatial distribution of visual extinction toward the Orion Molecular Ridge (see Section~3.1).  The lower white box
delineates the region containing BN/KL, the Orion Bar, and the low- and high-velocity outflows, while the upper white box delineates
the region containing OMC-2.  Both regions are excluded from the analysis presented
here.  All map offsets are relative to 
$\alpha\,=\,$05$^{\rm h}$35$^{\rm m}$14$^{\rm s}\!\!.$5,
$\delta\:=\,-$05\ddeg 22\amin 37\asec\ (J2000).}
\label{avmap}
\end{figure}

\clearpage

\begin{figure}[ht]
\centering
\vspace{-3mm}
\includegraphics[scale=0.95]{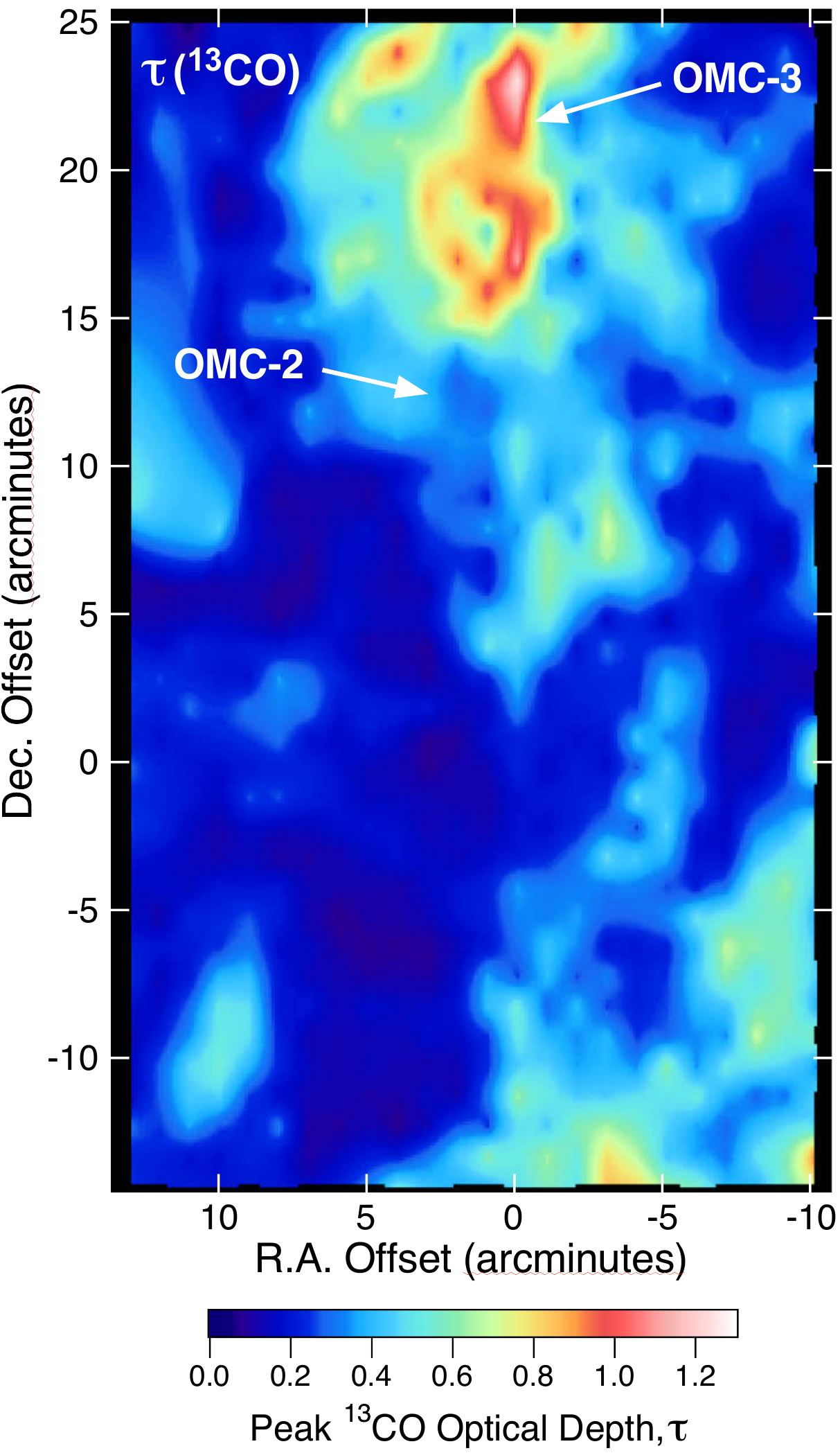}
\vspace{3.6mm}
\caption{Spatial distribution of peak Orion \ico\ optical depths based on the \nco\ and \ico\ measurements 
described in \S2.1 and the analysis presented in \S3.1.3.  All map offsets are relative to 
$\alpha\,=\,$05$^{\rm h}$35$^{\rm m}$14$^{\rm s}\!\!.$5,
$\delta\:=\,-$05\ddeg 22\amin 37\asec\ (J2000).}
\label{13cotaumap}
\end{figure}

\clearpage

\begin{figure}[ht]
\centering
\vspace{18mm}
\includegraphics[scale=0.83]{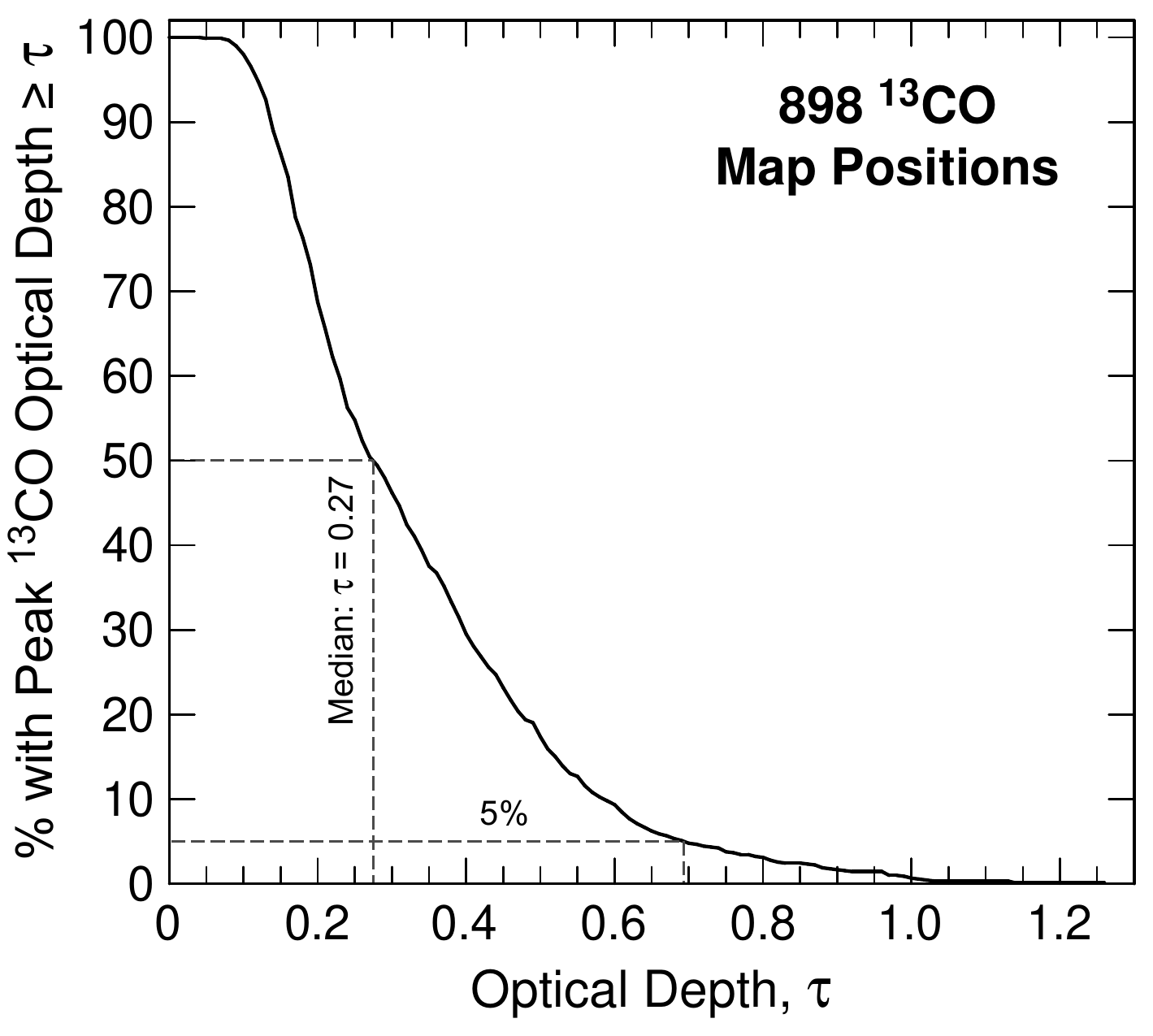}
\vspace{3.6mm}
\caption{Percentage of the 898 Orion \ico\ map positions within the water map area with peak \ico\ optical depths 
greater than the $x$-axis value.   The median peak \ico\ optical depth is 0.27 and more than 95\% of the peak \ico\ 
optical depths are less than 0.69.}
\label{13cotauplot}
\end{figure}

\clearpage

\begin{figure}[ht]
\centering
\vspace{2.5mm}
$\!\!$\includegraphics[scale=0.77]{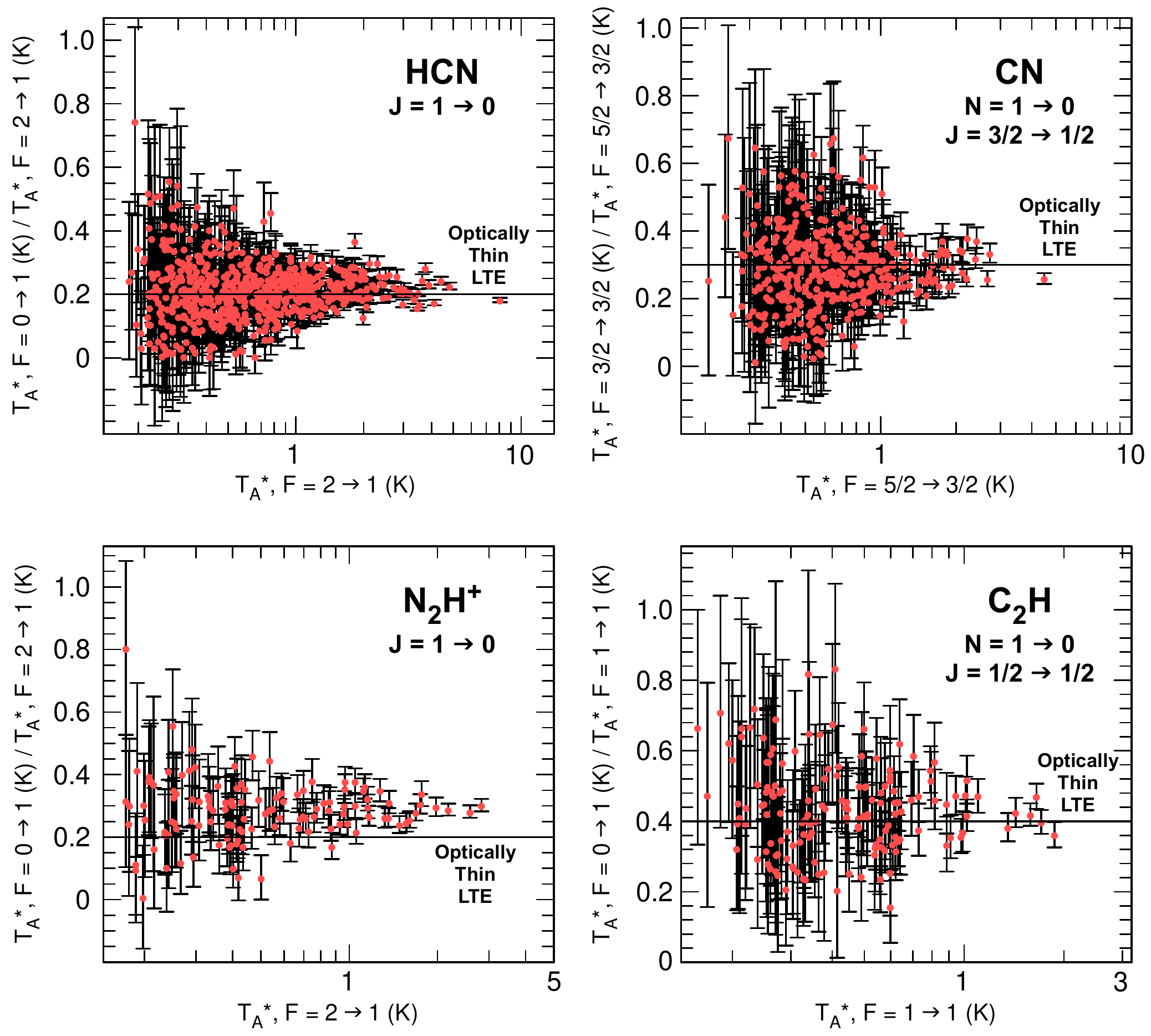}
\vspace{2.7mm}
\caption{Hyperfine intensity ratios toward Orion for HCN, CN, N$_2$H$^+$, and C$_2$H showing, in each panel, 
the ratio of the weakest hyperfine component to the strongest.  Such ratios are most sensitive to the optical depth for
each species.  Also shown are the ratios expected for optically thin LTE emission.  Although there is scatter and 
uncertainty in the ratio for the weaker intensity lines, as the lines become stronger, the ratio approaches the 
LTE optically thin value.  The slightly larger-than-LTE value for \nthp\ is discussed in the text.}
\label{hfsratios}
\end{figure}

\clearpage

\begin{figure}[ht]
\centering
\vspace{20mm}
$\!\!$\includegraphics[scale=0.785]{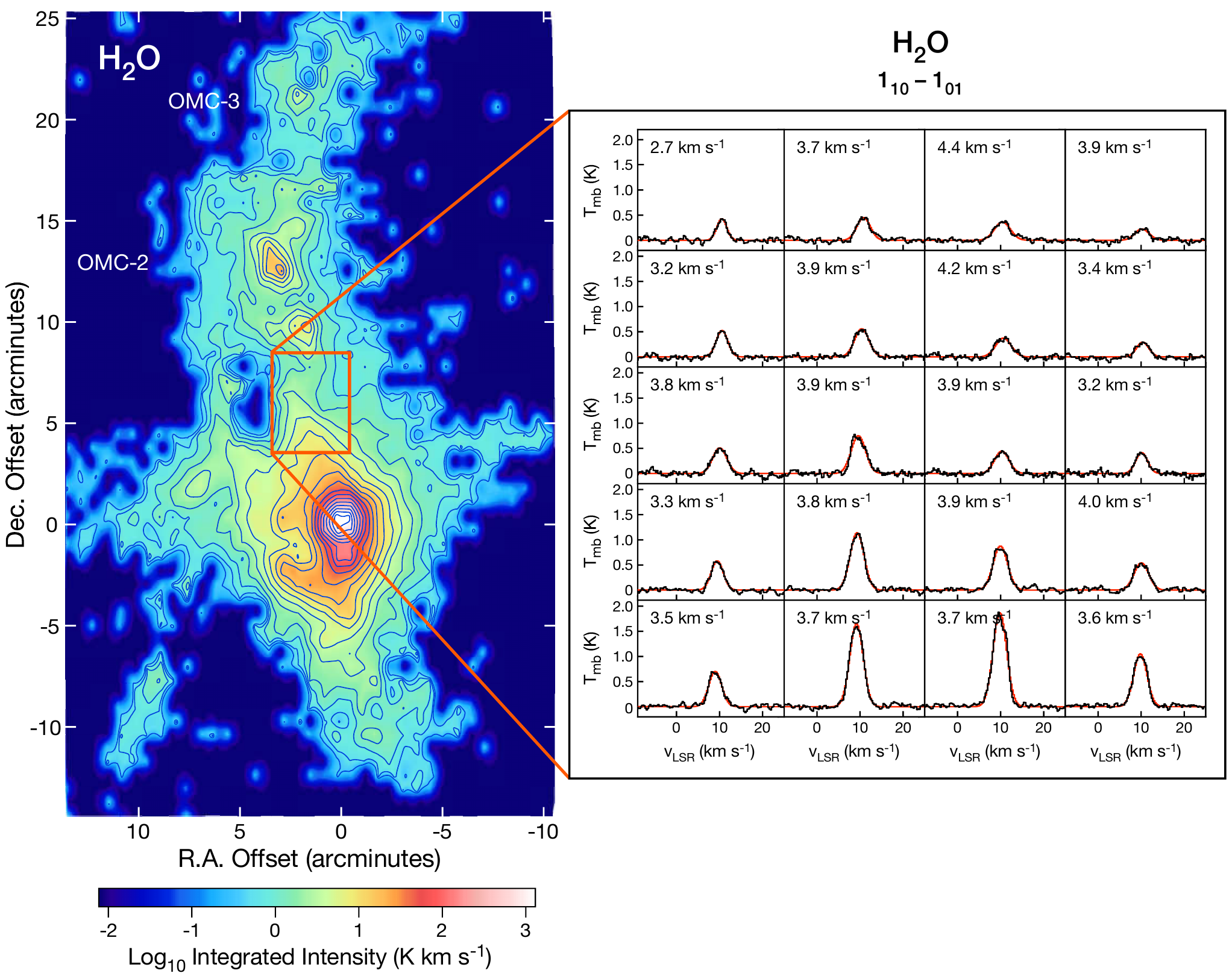}
\vspace{-0.7mm}
\caption{\water\ spectra obtained toward a portion of the Orion Ridge indicated by the boxed region on the left.  Each spectrum was
fitted with a single Gaussian line, in red, whose FWHM is given in the upper left of each panel.  In every case, a Gaussian
profile provides an excellent fit to the data.  This suggests that no significant frequency redistribution of escaping \water\ photons is 
occurring, which would result in a notable reduction of emission at line center and an enhancement of emission in the line wings.}
\label{h2ospectra}
\end{figure}

\clearpage

\begin{figure}[ht]
\centering
\vspace{-0.30in}
$\!$\includegraphics[scale=0.74]{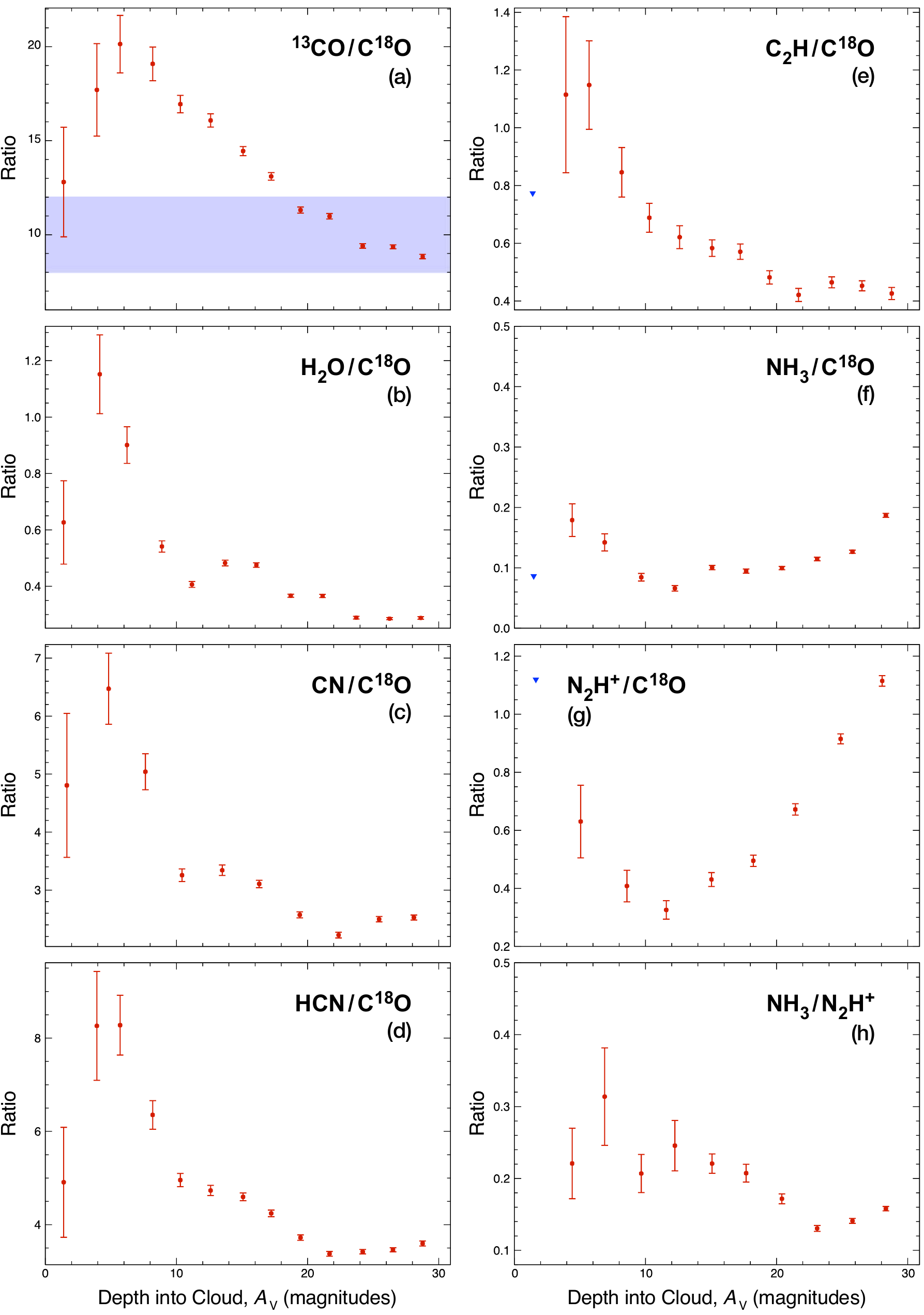}
\vspace{-2.0mm}
\caption{Plots of the ratios of the integrated intensities of the {\em Herschel}- and FCRAO-observed species 
versus depth into the Orion Molecular Ridge, measured in visual magnitudes, determined from the measured 
\ico\ column densities and the relations given in Eqns.~(1) and (2).  For clarity of presentation, 
the number of data points per plot has been reduced from 636 data points (for 0$\,\leq\,$\av$\,\leq\,$30) by 
averaging all data with a signal-to-noise ratio $\geq\:$3 within bins of \av.  The error bars for each point represent the
error-weighted mean and 1$\sigma$ uncertainty in the mean for the co-averaged points in each bin.
The purple region in panel (a) indicates the \ico\,/\,\cio\ ratio range expected
deep within the Ridge based on the $^{13}$C and $^{18}$O relative abundances measured toward Orion (see text).}
\label{ratiofig}
\end{figure}

\clearpage

\begin{figure}[ht]
\centering
\vspace{14mm}
$\!$\includegraphics[scale=0.67]{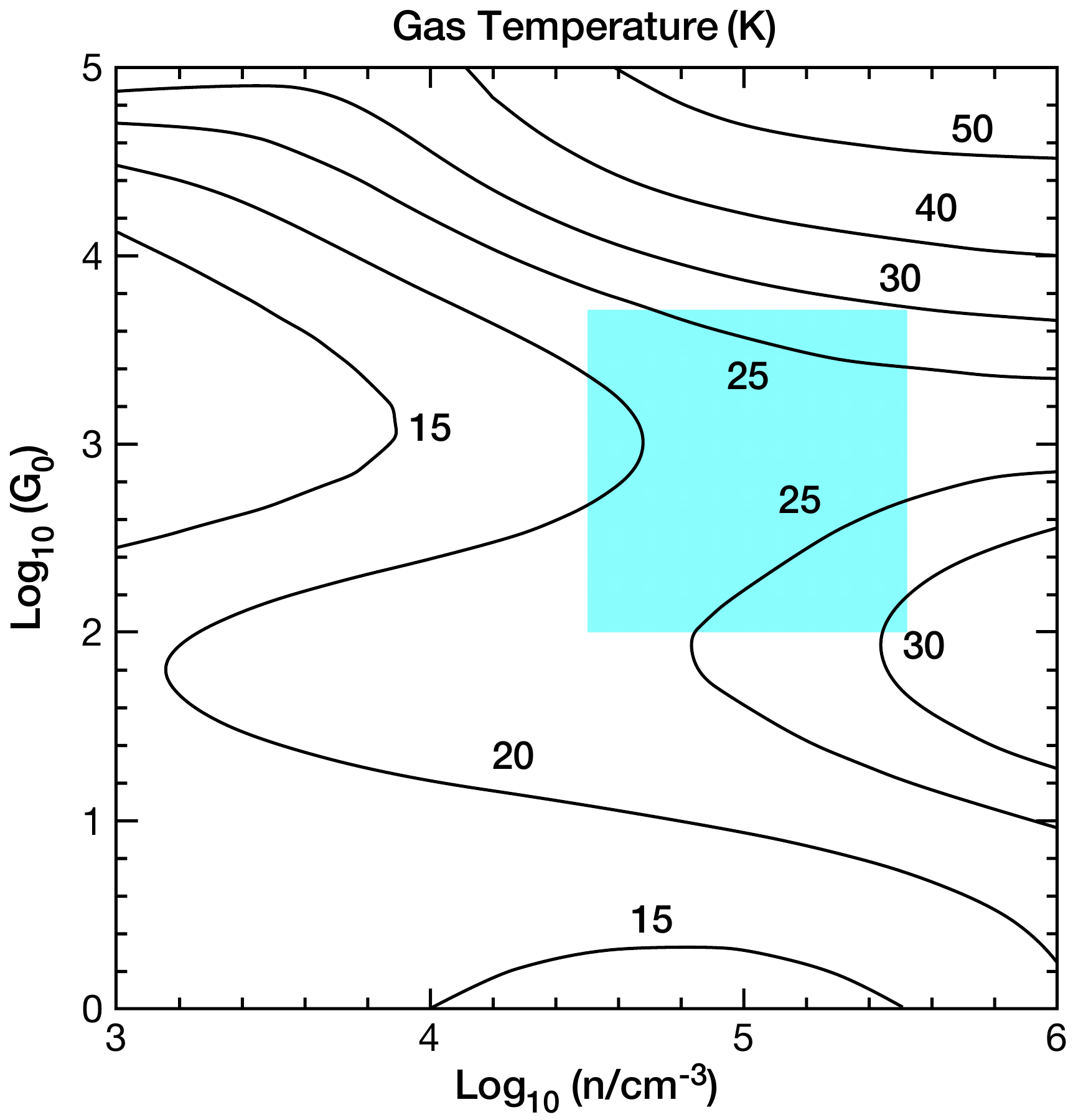}
\vspace{3.5mm}
\caption{Average gas temperature within the region of peak water abundance (after \citet{Hollenbach09}).  The blue box,
which has been added, shows the range of densities, $n$($= n_{\rm H} + $2$n_{H_2}$), and FUV field strengths, \go, 
appropriate to the Orion Molecular Ridge.  The average values for the Ridge are $n\:\sim\,$10$^5$~\cmc\ and
\go$\,\sim\:$500.
}
\label{gastemp}
\end{figure}

\clearpage

\begin{figure}[ht]
\centering
\vspace{7mm}
$\!$\includegraphics[scale=0.95]{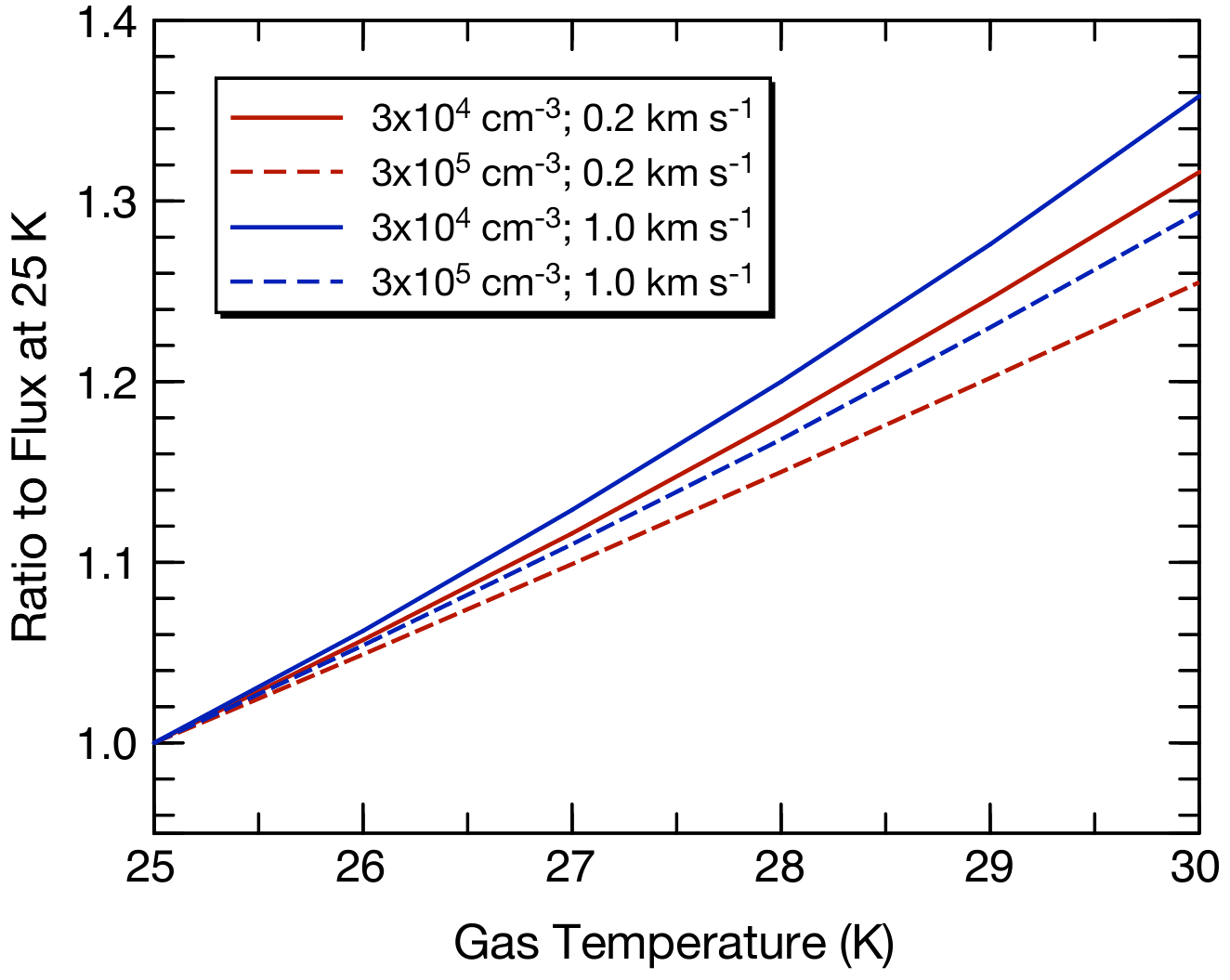}
\vspace{3.5mm}
\caption{Ratio of the \water\ \tro\ 557~GHz line flux calculated for a range of gas temperatures relative to that
at 25$\:$K.  \mh\ densities of 3\ti 10$^4\:$\cmc\ (solid lines) and 3\ti 10$^5\:$\cmc\ (dashed lines) 
and line widths per \av\ bin of 0.2\kms\ (red lines) and 1.0\kms\ (blue lines) are assumed.
An \mh\ column density per \av\ bin of 2.4\ti 10$^{21}\:$\cms\ is assumed, corresponding to an \av\ bin width of 2.5 magnitudes,
and an \water\ abundance of 10$^{-7}$.  (Note: The line flux ratios vary by less than 20 percent with a factor of 10 reduction
in the assumed \water\ abundance.)
}
\label{exciteration}
\end{figure}

%\clearpage

%\begin{figure}[h]
%\centering
%\vspace{0.60in}
%$\!\!\!$\includegraphics[scale=0.78]{H2O-13CO-C18O_PCA.pdf}
%\vspace{2.0mm}
%\caption{Results of the Principal Component Analysis for \water, \ico, and \cio\ for surface, intermediate, and deep
%layers within the Orion Molecular Ridge based on 101, 245, and 475 positions, respectively.  Principal Components 1 
%and 2 capture a cumulative fraction of 75\% of the total variance for the region between \av$\,=\,$0 and 4,
%86\% of the total variance for the region between \av$\,=\,$4 and 10, and 96\% of the total variance for the region 
%between \av$\,=\,$10 and 30.  Within the surface layer, the abundance of \water\ and \ico\ rise together as self-shielding
%becomes effective for both species; the abundance of \cio\ rises much more slowly in this region due to reduced
%self-shielding.  Within the intermediate and deep layers, \ico\ and \cio\ are fully-shielded and their abundances become highly
%correlated, as indicated by their closely aligned vectors in the middle and right panels; however, \water\ is no longer
%correlated with \ico\ and \cio\ within these layers.}
%\label{pdrpca}
%\end{figure}

\clearpage

\begin{figure}[ht]
\centering
\vspace{0.75in}
$\!\!$\includegraphics[scale=1.15]{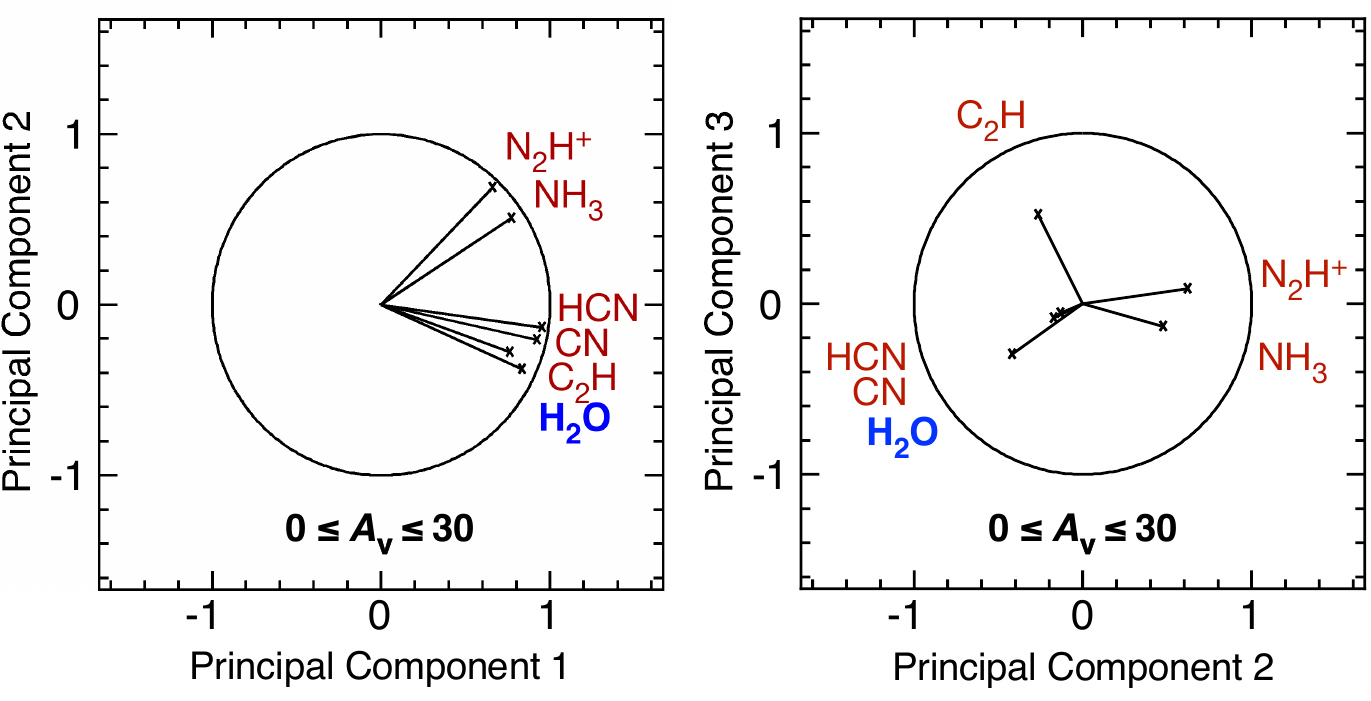}
\vspace{4.0mm}
\caption{Results of the Principal Component Analysis for \water, C$_2$H, CN, HCN, NH$_3$, and N$_2$H$^+$ for 820
Orion Molecular Ridge positions possessing cloud depths between \av$\,=\,$0 and 30. Principal Components 1 and 2 capture
a cumulative fraction of 87\% of the total variance while Principal Components 2\dash3 capture 94\% of the total variance.
Clearly, gas-phase \water\ is strongly correlated with HCN, CN, and C$_2$H while \water\ is largely uncorrelated with NH$_3$ and
N$_2$H$^+$, species that are expected to have their peak abundance deep within dense clouds.}
\label{pcafig}
\end{figure}

\clearpage

\begin{figure}[ht]
\centering
\vspace{-0.30in}
$\!$\includegraphics[scale=0.75]{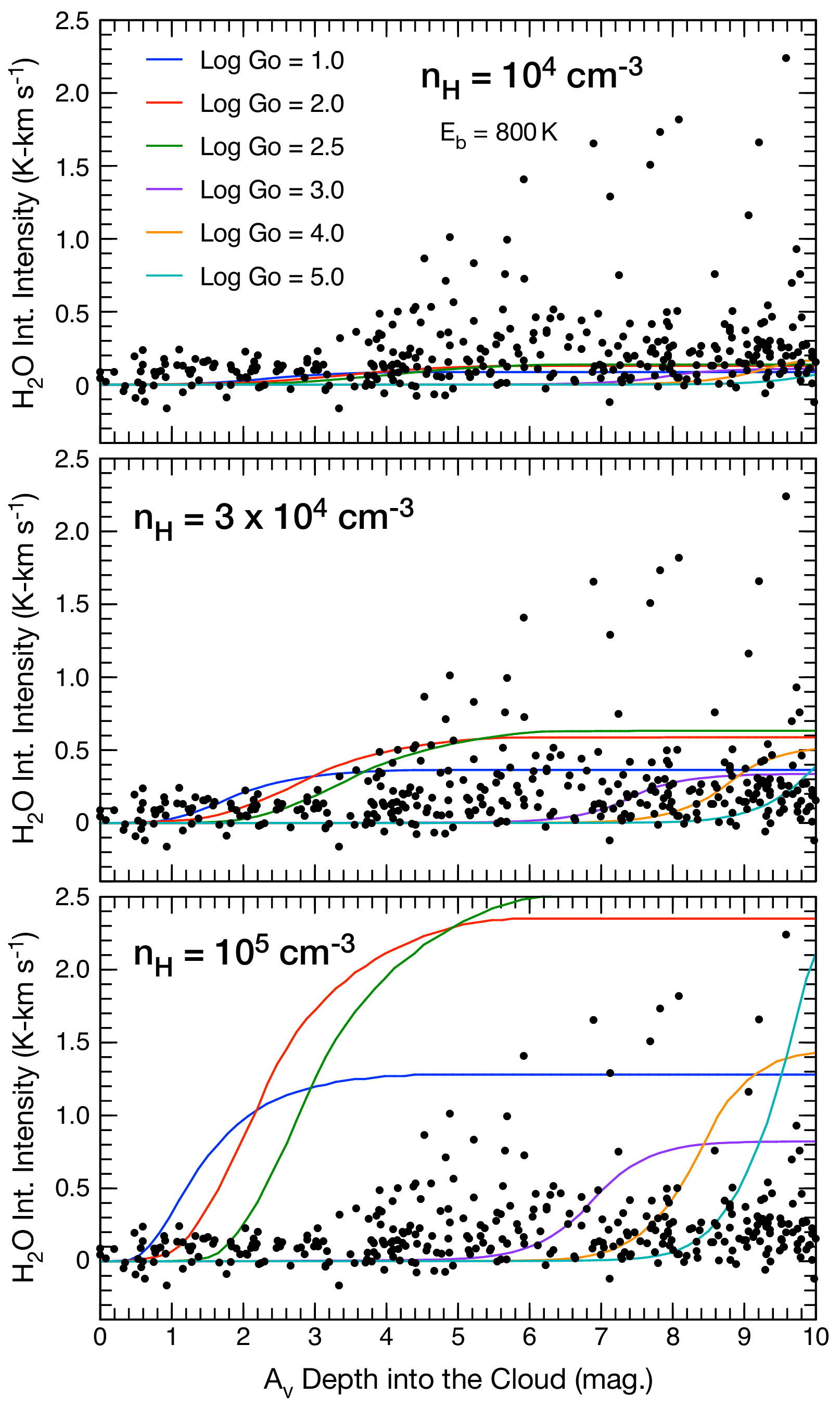}
%$\!$\includegraphics[scale=0.75]{H2OINTINT_vs_Av_vs_nH_vs_Go.pdf}
\vspace{-0.2mm}
\caption{Plots showing the gas-phase-\water\ integrated intensity versus \av\ for the 834 positions along the Orion Ridge (see
caption to Fig.~\ref{regridpos}).  Superposed on each panel are the model-predicted \water\ integrated intensities for 
an atomic oxygen-to-dust grain binding energy of 800$\:$K \citep{Tielens82}
and various values of \go\ and gas densities of 10$^4$~\cmc\ ({\em top panel}), 3\ti 10$^4$~\cmc\ ({\em middle panel}), 
and 10$^5$~\cmc\ ({\em bottom panel}).}
\label{Eb800_Plot}
\end{figure}

\clearpage

\begin{figure}[ht]
\centering
\vspace{-0.30in}
$\!$\includegraphics[scale=0.75]{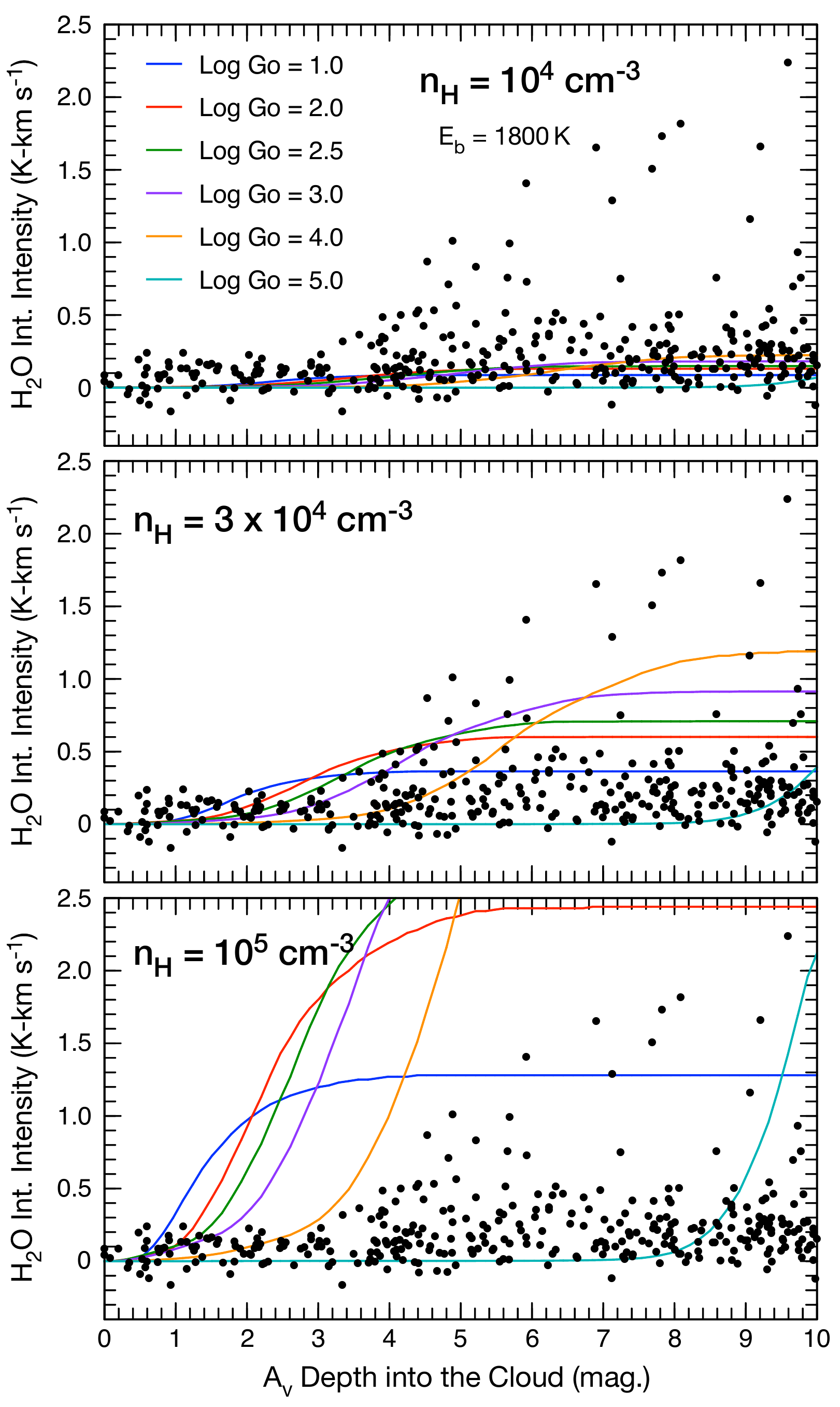}
%$\!$\includegraphics[scale=0.75]{H2OINTINT_vs_Av_vs_nH_vs_Go.pdf}
\vspace{-0.2mm}
\caption{Plots showing the gas-phase-\water\ integrated intensity versus \av\ for the 834 positions along the Orion Ridge (see
caption to Fig.~\ref{regridpos}).  Superposed on each panel are the model-predicted \water\ integrated intensities for 
an atomic oxygen-to-dust grain binding energy of 1800$\:$K \citep{He15}
and various values of \go\ and gas densities of 10$^4$~\cmc\ ({\em top panel}), 3\ti 10$^4$~\cmc\ ({\em middle panel}), 
and 10$^5$~\cmc\ ({\em bottom panel}).}
\label{Eb1800_Plot}
\end{figure}

\clearpage

\begin{figure}[ht]
\centering
\vspace{1.5in}
$\!\!$\includegraphics[scale=0.93]{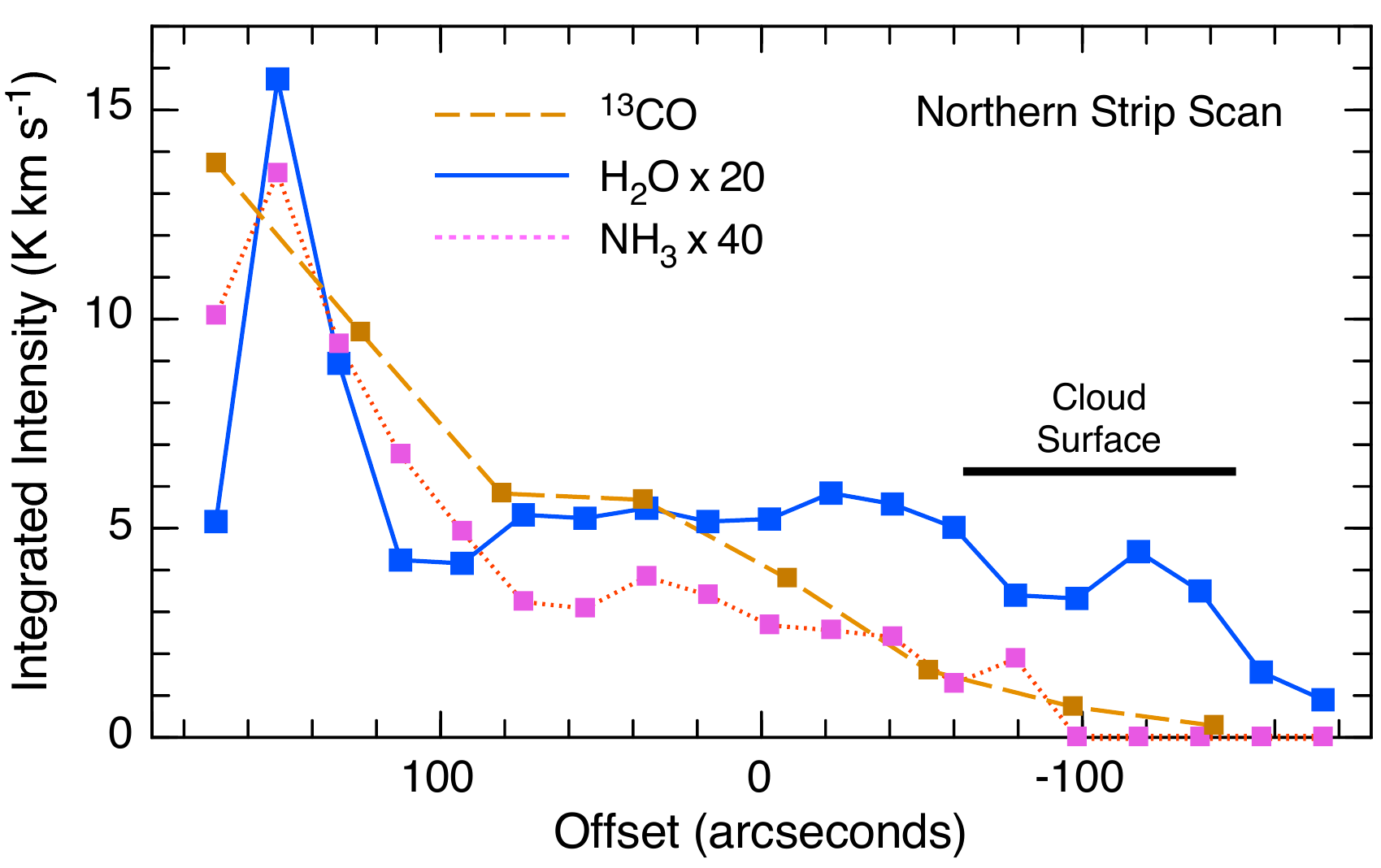}
\vspace{4.3mm}
\caption{Integrated intensity of \ico, \water, and NH$_3$ as a function of offset from the Cepheus B northern strip scan (0,0)
position, i.e., R.A. 22$^{\rm h}$ 57$^{\rm m}$ 16$^{\rm s}$, Dec. $+$62\ddeg\ 35\amin\ 45\asec\ (J2000). The \ico\
strip scan was obtained using FCRAO, while the \water\ and NH$_3$ strip scans were obtained using {\em Herschel}. 
In order to better illustrate the trends, the \water\ integrated intensities have been multiplied by 20, and the NH$_3$
integrated intensities have been multiplied by 40.}
\label{stripscannorth}
\end{figure}

\clearpage

\begin{figure}[ht]
\centering
\vspace{1.5in}
$\!\!$\includegraphics[scale=0.93]{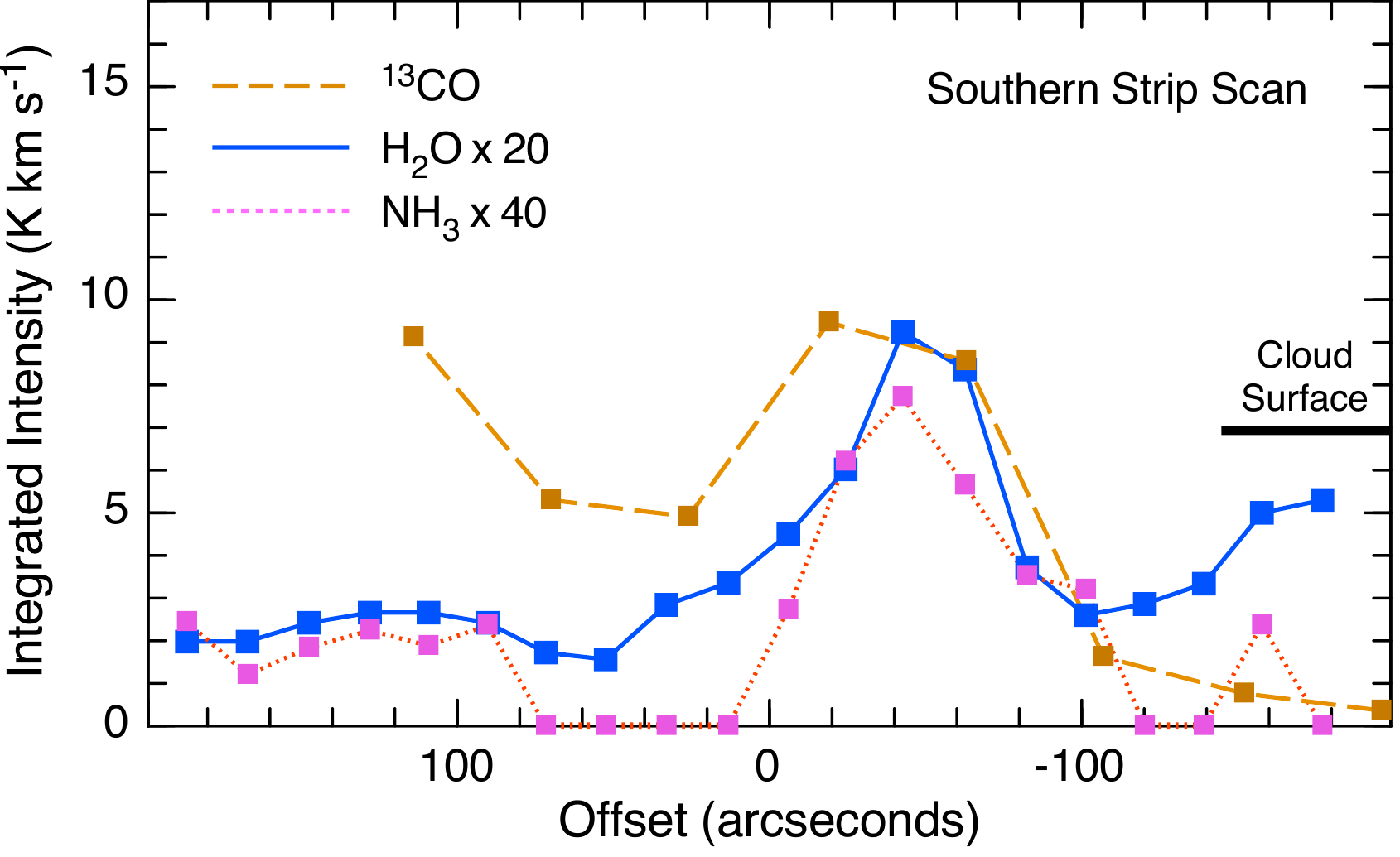}
\vspace{4.3mm}
\caption{Integrated intensity of \ico, \water, and NH$_3$ as a function of offset from the Cepheus B southern strip scan (0,0)
position, i.e., R.A. 22$^{\rm h}$ 57$^{\rm m}$ 24$^{\rm s}$, Dec. $+$62\ddeg\ 34\amin\ 45\asec\ (J2000). The \ico\
strip scan was obtained using FCRAO, while the \water\ and NH$_3$ strip scans were obtained using {\em Herschel}. 
In order to better illustrate the trends, the \water\ integrated intensities have been multiplied by 20, and the NH$_3$
integrated intensities have been multiplied by 40.}
\label{stripscansouth}
\end{figure}

\clearpage

\begin{figure}[ht]
\centering
\vspace{1.5in}
$\!\!\!\!\!$\includegraphics[scale=0.74]{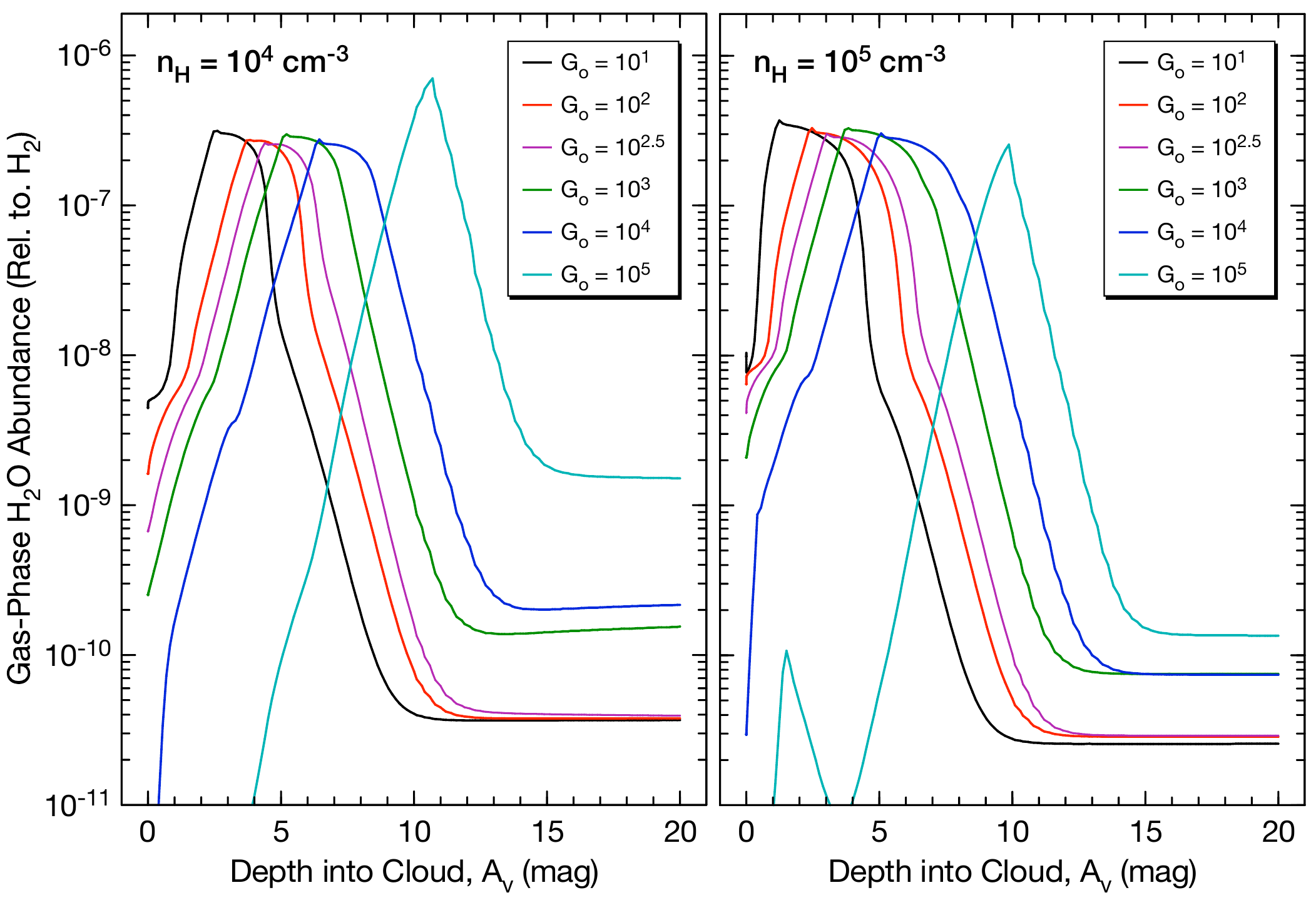}
\vspace{2.0mm}
\caption{Predicted gas-phase \water\ abundance vs.~depth into clouds, measured in magnitudes of visual extinction, \av,
and FUV flux, \go, for clouds of gas density 10$^4$~\cmc\ and 10$^5$~\cmc.  An atomic oxygen binding energy of 1800$\:$K
is assumed (see text).
%The curves are labelled and color coded by log\,\go.
}
\label{h2oabund}
\end{figure}

\clearpage

\begin{figure}[ht]
\centering
\vspace{1.0in}
$\!\!$\includegraphics[scale=0.85]{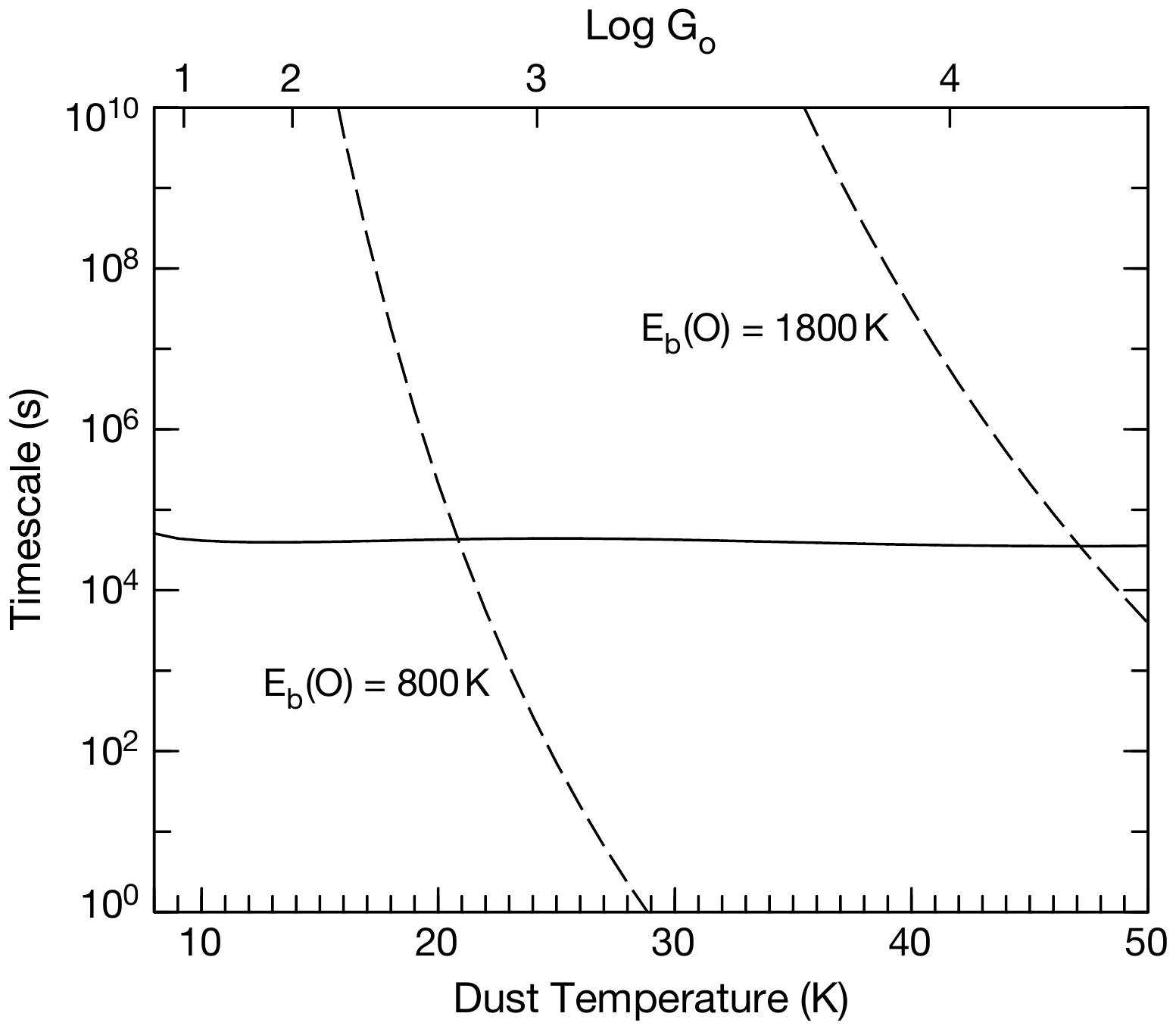}
\vspace{4.3mm}
\caption{Timescales relevant to OH formation on grains.  The dashed curves show the timescale for
thermal desorption of an O atom on a grain for both the old (800$\:$K) and new (1800$\:$K) value of the O binding energy.  
The solid curve is the timescale for formation of an OH molecule given an O atom adsorbed to the surface, which 
is dominated by the time it takes for an H atom to arrive on
the grain surface (see text).  The relation between the dust grain temperature and \go\ is that provided in \citet{Hollenbach09}.
Once the H atom has adsorbed onto the surface, its diffusion is so rapid that the reaction with O is essentially instantaneous 
(compared to the H-accretion timescale).  
At dust temperatures where the solid line is below the dashed line, O atoms remain
on grain surfaces sufficiently long to combine with an H atom to form OH (and, ultimately, \water).
}
\label{lifetime_plot}
\end{figure}

\end{document}